\documentclass[useAMS,usenatbib,letter]{mn2e}

\usepackage{amsmath} 
\usepackage{subfigure} 
\usepackage{epsfig}
\usepackage{appendix}
\usepackage{xspace}
\usepackage{multirow}
\usepackage[T1]{fontenc}
\usepackage{amssymb}
\usepackage{url}
\usepackage[colorlinks=false,linkcolor=black, citecolor=green, linktoc=page, urlcolor=cyan]{hyperref}
\usepackage[mathlines]{lineno}

\def\reff@jnl#1{{\rm#1\/}}
\def\aj{\reff@jnl{AJ}}         
\def\araa{\reff@jnl{ARA\&A}}      
\def\apj{\reff@jnl{ApJ}}        
\def\apjl{\reff@jnl{ApJ}}        
\def\apjs{\reff@jnl{ApJS}}       
\def\aap{\reff@jnl{A\&A}}        
\def\aapr{\reff@jnl{A\&A~Rev.}}     
\def\aaps{\reff@jnl{A\&AS}}       
\def\mnras{\reff@jnl{MNRAS}}      
\def\physrep{\reff@jnl{Physics Reports}}
\def\prd{\reff@jnl{Phys.Rev.D}}     
\def\prl{\reff@jnl{Phys.Rev.Lett}}   
\def\pasp{\reff@jnl{PASP}}       
\def\pasj{\reff@jnl{PASJ}}       
\def\jcap{\reff@jnl{JCAP}}   
\def\nat{\reff@jnl{Nature}}       

\def\Sref#1{Section~\ref{#1}\xspace}
\def\srefa#1{Sections~\ref{#1}\xspace}
\def\srefb#1{\ref{#1}\xspace}
\def\Fref#1{Figure~\ref{#1}\xspace}
\def\frefa#1{Figures~\ref{#1}\xspace}

\def\Tref#1{Table~\ref{#1}\xspace}

\def\Eref#1{Equation~\ref{#1}\xspace}

\def\Aref#1{Appendix~\ref{#1}\xspace}
\def\eg{{\it e.g.}}
\def\ie{{\it i.e.}}
\def\etc{{\it etc.}}

\def\phosim{{\sc PhoSim}\xspace}
\def\opsim{{\sc OpSim}\xspace}
\def\catsim{{\sc CatSim}\xspace}
\def\imcat{{\sc imcat}\xspace}

\def\sigmag{\hat{\sigma}_{\gamma}}
\def\neff{n_{\rm eff}}
\def\neffh{n_{\rm eff}^{*}}

\usepackage[usenames]{color}
\usepackage{graphicx}

\def\kipac{KIPAC, Stanford University, 452 Lomita Mall, 
Stanford, CA 94309, USA}

\def\purdue{Department of Physics, Purdue University, 
 West Lafayette, IN 47907, USA}
\def\uw{Department of Astronomy, University of Washington, 
Seattle, WA 98195}

\def\upenn{Department of Physics and Astronomy, University of Pennsylvania, 
209 South 33rd Street, Philadelphia, PA 19104, USA}
\def\irvine{Department of Physics and Astronomy, University of California, 
Irvine, 92697, USA}

\title[$\neff$ for the LSST Project]
{The Effective Number Density of Galaxies for Weak Lensing Measurements in the LSST Project}

\author[C.~Chang et al.]
{C.~Chang,$^{1}$\thanks{E-mail: chihway@slac.stanford.edu}
M.~Jarvis,$^{2}$
B.~Jain, $^{2}$
S.~M.~Kahn,$^{1}$
D.~Kirkby$^{3}$
\newauthor 
A.~Connolly,$^{4}$
S.~Krughoff,$^{4}$
E.~Peng,$^{5}$  
J.~R.~Peterson$^{5}$ \\
$^{1}$\kipac\\
$^{2}$\upenn\\
$^{5}$\purdue \\
$^{3}$\irvine\\
$^{4}$\uw
}

\begin{document}
\topmargin=-0.5in

\date{Accepted, Received; in original form }

\pagerange{\pageref{firstpage}--\pageref{lastpage}} \pubyear{2011}

\maketitle

\label{firstpage}


\begin{abstract}
Future weak lensing surveys potentially hold the highest statistical power for constraining cosmological parameters compared to 
other cosmological probes. The statistical power of a weak lensing survey is determined by the sky coverage, the inverse of the noise 
in shear measurements, and the galaxy number density. The combination of the latter two factors is often expressed in terms of $\neff$ 
-- the ``effective number density of galaxies used for weak lensing measurements''. In this work, we estimate $\neff$ for the Large 
Synoptic Survey Telescope (LSST) project, the most powerful ground-based lensing survey planned for the next two decades. We 
investigate how the following factors affect the resulting $\neff$ of the survey with detailed simulations: (1) survey time, (2) shear 
measurement algorithm, (3) algorithm for combining multiple exposures, (4) inclusion of data from multiple filter bands, (5) redshift 
distribution of the galaxies, and (6) masking and blending. For the first time, we quantify in a general weak lensing analysis pipeline 
the sensitivity of $\neff$ to the above factors. 

We find that with current weak lensing algorithms, expected distributions of observing parameters, and all lensing data ($r$- and 
$i$-band, covering 18,000 degree$^{2}$ of sky) for LSST, $\neff \approx37$ arcmin$^{-2}$ before considering blending and masking, 
$\neff \approx31$ arcmin$^{-2}$ when rejecting seriously blended galaxies and $\neff \approx26$ arcmin$^{-2}$ when considering 
an additional 15\% loss of galaxies due to masking. With future improvements in weak lensing algorithms, these values could be 
expected to increase by up to 20\%.
Throughout the paper, we also stress the ways in which $\neff$ depends on 
our ability to understand and control systematic effects in the measurements.

\end{abstract}



\section{Introduction}
Weak lensing is one of the most powerful tools for probing the dark matter distribution 
in our Universe and constraining dark energy parameters \citep{2012arXiv1201.2434W}. Gravitational fields 
due to the large scale matter distribution perturb light rays traveling from distant galaxies, causing the 
observed galaxy shapes to be slightly distorted compared to their true shapes. Since the original 
shapes are unknown, these weak distortions cannot be discovered by observations of individual galaxies. They can 
only be inferred via statistical approaches, \eg, correlations of galaxy shape parameters as a function of 
angular separation \citep[see, \eg,][]{2001PhR...340..291B}. 

The ultimate statistical power for deriving cosmological parameters with weak lensing depends on the total 
sky coverage and the number density of galaxies with accurate shear measurements in a survey. One reduces the 
statistical uncertainties from weak lensing by surveying wider and deeper fields. This has been the driver behind all 
ongoing and future weak lensing surveys [\eg, The Kilo Degree Survey\footnote{\url{http://www.astro-wise.org/projects/KIDS/}} 
(KiDS), the Hyper Suprime-Cam Survey\footnote{\url{http://www.astro.princeton.edu/~rhl/HSC/}} (HSC), the Dark Energy 
Survey\footnote{\url{http://www.darkenergysurvey.org/}} (DES), the Large Synoptic Survey 
Telescope\footnote{\url{http://www.lsst.org/lsst/}} (LSST), the Euclid 
mission\footnote{\url{http://sci.esa.int/science-e/www/object/index.cfm?fobjectid=42266}}, and the Wide-Field 
Infrared Survey Telescope\footnote{\url{http://wfirst.gsfc.nasa.gov/}} (WFIRST)]. 

Using the weak lensing Fisher-matrix calculation introduced by \citet[][hereafter A06]{2006astro.ph..9591A}, 
the statistical uncertainty of a weak lensing survey is determined by the combined quantity $f_{sky}^{-0.5} \sigmag^2$, 
where $f_{sky}$ is the fraction of sky covered by the survey and $\sigmag$ is the uncertainty on the mean shear in 
unit area. The sky coverage $f_{sky}$ is limited by the accessible sky from a ground-based telescope. $\sigmag$, 
however, cannot be straightforwardly calculated, since it contains not only the designed survey depth (the number of 
galaxies we see given the design of the survey), but also the performance of the entire analysis pipeline one uses to 
measure these distortions (the amount of the information we can actually extract from each galaxy given the measurement 
methods). 

In A06, the question of calculating $\sigmag$ is rephrased in terms of calculating $\neff$, the effective number 
density of galaxies used for weak lensing measurements. A06 defined $\neff$ to be \textit{the number density of 
perfectly measured galaxies that would contribute the same amount of shear noise as the 
(imperfectly) measured ensemble of galaxies}. The standard deviation in each component of the 
ellipticity for the perfectly measured galaxy population (commonly known as ``shape noise''), 
$\sigma_{SN}$, is assumed to be fixed at 0.25 in A06. The term ``noise'' refers to the fact that the 
intrinsic galaxy shapes introduce uncertainty in the shear inferred from these galaxies. Most studies 
to date have adopted the formulae and model parameters in A06 to estimate the performance 
of future weak lensing surveys. However, the ways in which $\neff$ is quoted in the literature are often 
inconsistent, causing confusion in the field. The main goal of this paper is to clearly define $\neff$, 
estimate $\neff$ for LSST, and quantitatively evaluate the sensitivity of $\neff$ to different assumptions about the 
survey plan and analysis pipeline. Our methodology for calculating $\neff$ is general and can be 
applied to other surveys given the basic survey and analysis information.

In \Sref{sec:overview}, we first review briefly the weak lensing 
notation and definition of $\neff$. We also discuss the relevant information about the dataset and the 
analysis pipeline required to calculate $\neff$. We introduce the input galaxy catalog 
and the $\sigma_{SN}$ value for this work in \Sref{sec:sigmaSN}. Then we show step by step in 
\Sref{sec:sigmaM} how the shear measurement noise is estimated depending on different analysis 
methods and galaxy selection. We calculate in \Sref{sec:neff} the $\neff$ values for LSST under different 
scenarios and discuss in depth the different factors that affect $\neff$. Finally, we estimate in 
\Sref{sec:practicalities} how $\neff$ degrades when practical effects such as masking and blending are 
introduced. The same calculation is then applied to an existing survey in \Sref{sec:cfhtls} to demonstrate 
the generality of our approach. We summarize our results in \Sref{sec:conclusion}.

\section{Overview of the problem}
\label{sec:overview}

\subsection{Weak lensing notation}
\label{sec:notation}

Throughout the paper, we measure object shapes using the 2-component ellipticity spinor:
\begin{linenomath*}\begin{equation} 
  \boldsymbol{\epsilon}=\epsilon_{1}+i\epsilon_{2}\;,
\end{equation} \end{linenomath*}
\noindent where
\begin{equation*}
  \epsilon_{1} =\frac{I_{11}-I_{22}}{I_{11}+I_{22}+2\sqrt{I_{11}I_{22}-I_{12}^{2}}} \; , 
\end{equation*}
\begin{linenomath*}\begin{equation} 
  \epsilon_{2} =\frac{2I_{12}}{I_{11}+I_{22}+2\sqrt{I_{11}I_{22}-I_{12}^{2}}} \;.
  \label{eq:ellipticity}
\end{equation} \end{linenomath*}
\noindent The $I_{ij}$ are the normalized moments of the object's light intensity profile 
$I(x_{1},x_{2})$: 
\begin{linenomath*}\begin{equation}  
  I_{ij}=\frac{\int \int dx_{1}dx_{2}I(x_{1},x_{2})x_{i}x_{j}}
  {\int \int dx_{1}dx_{2}I(x_{1},x_{2})}, \; i,j=1,2  \;.
\label{eq:moments}
\end{equation} \end{linenomath*}
Under this definition, the measured ellipticity $\boldsymbol{\epsilon}$ changes accordingly in the presence of 
shear $\boldsymbol{\gamma}=\gamma_{1}+i \gamma_{2}$ and convergence $\kappa$ \citep[see, \eg, ][]{2001PhR...340..291B}: 
\begin{linenomath*}\begin{equation} 
\boldsymbol{\epsilon}= 
\begin{cases} \; (\boldsymbol{\epsilon^{s}}+\boldsymbol{g})(1+\boldsymbol{g}^{*}\boldsymbol{\epsilon^{s}})^{-1} \;  &,|\boldsymbol{g}| \leq 1 \\
                          \; (1+\boldsymbol{g}\boldsymbol{\epsilon^{s,*}})(\boldsymbol{\epsilon^{s,*}}+\boldsymbol{g^{*}})^{-1} \; &,|\boldsymbol{g}| > 1 \\
\end{cases}\;\;;
\label{eq:shear_e}
\end{equation} \end{linenomath*}
where $\boldsymbol{\epsilon^{s}}=\epsilon^{s}_{1}+i \epsilon^{s}_{2}$ refers to the intrinsic ellipticity of the galaxy before shearing, the 
asterisk denotes the complex conjugate, and $\boldsymbol{g}=g_{1}+i g_{2}$ is the ``reduced shear'' defined by
\begin{linenomath*}\begin{equation} 
\boldsymbol{g}=\frac{\boldsymbol{\gamma}}{1- \kappa}\;\;.
\end{equation} \end{linenomath*}
Here we have adopted the approximation that $\boldsymbol{\gamma} \approx \boldsymbol{g}$ in the limit of weak lensing where 
$\kappa \ll 1$.

In the presence of noise, a weighting function $W(x_{1},x_{2})$ is included in the integrands in \Eref{eq:moments} 
to reduce the fluctuations in ellipticity measurements. The width of $W(x_{1},x_{2})$ is approximately the size of the 
observed object -- this yields the maximum signal-to-noise ratio for each individual object. Due to imperfect 
point-spread-function (PSF) models and this weighting function $W(x_{1},x_{2})$, \Eref{eq:shear_e} is no longer exact. 
As a result, the ``shear responsivity'' \citep{1997ApJ...475...20L} is introduced to correct for this effect. 
In practice, due to the existence of noise-induced biases \citep{2012MNRAS.425.1951R, 2012MNRAS.424.2757M}, 
the shear measurements need to be further calibrated from 
simulations \citep{2006MNRAS.368.1323H, 2007MNRAS.376...13M, 2010MNRAS.405.2044B}. 

\subsection{Relation between shear noise and $\neff$}
\label{sec:clarification}

As mentioned earlier, the relevant quantity in measuring the statistical power of a survey is the uncertainty on the mean 
shear per unit area, or $\sigmag$ (given fixed $f_{sky}$). 
For each galaxy, since the shear noise results from the intrinsic shape noise as well as measurement noise, we 
can write,
\begin{linenomath*}\begin{equation} 
\sigma_{\gamma,i}^2=\sigma_{SN}^2+\sigma_{m,i}^2\;\;,
\label{eq:shear_breakdown}
\end{equation} \end{linenomath*}
where we have assumed that shape noise is uncorrelated with measurement noise.
The subscript $i$ indicates this is the shear noise for the $i$th galaxy and the subscript $m$ 
refers to the measurement noise. $\sigma$ indicates the Root-Mean-Square (RMS) of the distribution. 
Note that measurement noise depends on the galaxy's shape, size and brightness, while shape 
noise is usually taken to be constant for the entire galaxy sample (however, see discussion in 
\Sref{sec:sigmaSN}). 

If we assume the mean shear estimation $\hat{\gamma}$ is calculated by the weighted mean of the shear over the 
entire sample, where the weight is just the inverse variance in each measurement, then we have: 
\begin{linenomath*}\begin{equation} 
\hat{\gamma}=\frac{\Sigma_{i}^{N}\frac{\gamma_{i}}
{\sigma^2_{\gamma,i}}}{\Sigma_{i}^{N}\frac{1}{\sigma^2_{\gamma,i}}}.
\label{eq:mean_shear}
\end{equation} \end{linenomath*}
$\sigmag$ is equal to the survey area times the variance in $\hat{\gamma}$, which is derived from the 
shear noise in individual galaxies, $\sigma_{\gamma,i}$: 
\begin{linenomath*}\begin{equation} 
\sigmag^{2}=\Omega {\rm Var}(\hat{\gamma}) = \Omega \left [\Sigma_{i}^{N}\frac{1}{\sigma_{\gamma,i}^2}\right ]^{-1}
\equiv \frac{\Omega \sigma_{SN}^2}{N_{\rm eff}} \equiv \frac{\sigma_{SN}^2}{\neff},
\label{eq:sum_shear}
\end{equation} \end{linenomath*}
where $N$ is the total number of galaxies used in the lensing analysis and $\Omega$ is the total sky coverage. 
The last two factors in \Eref{eq:sum_shear} provide the operational definition of $\neff$, where we have defined $N_{\rm eff}$ 
to be the effective number of weak lensing galaxies corresponding to this galaxy sample and 
$\neff = N_{\rm eff}/\Omega$. Rearranging the terms in \Eref{eq:sum_shear} and using 
\Eref{eq:shear_breakdown} leads to the following relation: 
\begin{linenomath*}\begin{equation} 
\neff=\frac{\sigma_{SN}^{2}}{\sigmag^{2}}=\frac{1}{\Omega}\Sigma_{i}^{N}\frac{\sigma_{SN}^2}{\sigma_{\gamma,i}^2}
=\frac{1}{\Omega}\Sigma_{i}^{N}\frac{\sigma_{SN}^2}{\sigma_{SN}^2+\sigma_{m,i}^2}\;\;.
\label{eq:neff}
\end{equation} \end{linenomath*}
Recall the measured weak lensing power spectrum can be written as \citep{2008MNRAS.391..228A}
\begin{linenomath*}\begin{equation} 
C_{ij}^{\gamma}(\ell)=P_{ij}^{\gamma}(\ell)+\delta_{ij}\sigmag^{2}+ C_{ij}^{\gamma, \rm sys}(\ell)\;\;, 
\end{equation} \end{linenomath*}
where $i$ and $j$ denote two redshift bins, $C_{ij}^{\gamma}(\ell)$ is the measured lensing power spectrum, 
$P_{ij}^{\gamma}(\ell)$ is the true lensing power spectrum, $\delta_{ij}$ is the Kronecker delta function and 
$C_{ij}^{\gamma, \rm sys}(\ell)$ is the systematic error in the shear power spectrum measurement. The uncertainty 
in the measured weak lensing power spectrum can be written as
\begin{linenomath*}\begin{equation} 
\Delta P_{ij}^{\gamma}(\ell)=\sqrt{\frac{2}{(2 \ell +1)f_{sky}}}\left[ P_{ij}^{\gamma}(\ell)+\delta_{ij}\sigmag^{2} 
+ C_{ij}^{\gamma, \rm sys}(\ell) \right].
\label{eq:error_ps}
\end{equation} \end{linenomath*}
This clearly displays what was mentioned earlier -- the statistical uncertainty in cosmic shear measurements 
(the second term in the square brackets) is determined by the factor $f_{sky}^{-0.5}\sigmag^{2}$, or equivalently, 
$f_{sky}^{-0.5}\sigma_{SN}^{2}\neff^{-1}$.

From \Eref{eq:neff}, we observe that calculating $\neff$ for LSST involves a combination of considerations. 
First, we need an understanding of the intrinsic distribution of galaxies in the multi-dimensional 
space (\eg, size, magnitude, redshift, shape, \etc), given the depth of the LSST dataset. Second, we need to 
understand the expected shear measurement error for each galaxy, which depends on the characteristics 
of the galaxy, the measurement algorithm, how multiple measurements of the same galaxy are combined, and 
considerations of systematic errors in the measurement. We address the first part of the problem (the intrinsic 
galaxy distribution) in \Sref{sec:sigmaSN} and the second part (the shear measurement error) in \Sref{sec:sigmaM}.
It is important to realize that even for the same dataset, it is possible to get different $\neff$ values depending 
on the different choices one makes with the analysis pipeline. Thus one needs to be careful when quoting or 
comparing these numbers, to give enough information on the assumptions involved.

We also note that in \Eref{eq:error_ps} we have intentionally separated the systematic errors from the statistical 
errors in the $\neff$ calculation for simplicity. However, as we discuss in \Sref{sec:sigmaM}, $\neff$ depends 
on several factors in the analysis pipeline that are set by requirements on systematic errors in shear 
measurements. As a result, $\neff$ can be coupled with the systematic errors in an indirect way. The exact 
tradeoff between systematic and statistical errors for cosmic shear measurements, and the effect on $\neff$ is 
algorithm-dependent and beyond the scope of this paper.

\section{Shape noise and the intrinsic galaxy distribution}
\label{sec:sigmaSN}

To start, we need a realistic galaxy catalog that contains the primary characteristics (redshift, size, magnitude 
and shape) of the galaxies expected to be seen in a 10-year LSST weak lensing dataset. The LSST weak 
lensing survey is expected to image 18,000 square degrees \citep{SRD}
of the sky in six filter bands 
($ugrizy$) to a median redshift of $\sim$1.2 and depth of $r\sim27.5$ and $i\sim26.8$\footnote{This magnitude 
limit is defined as the $r$-band AB magnitude at 5$\sigma$ for a point source.} \citep{2008arXiv0805.2366I}. 
For this study, we use a typical simulated galaxy catalogue generated by the LSST Catalog Simulator
\citep[][\catsim]{A13}. 

We briefly describe here the key steps and references for creating the galaxy catalog to help readers 
understand the results of this work. 
First, galaxies in the simulated galaxy catalog from \citet{2006MNRAS.366..499D} are matched with the 
dark matter peaks in the Millennium Simulation 
\citep{2005Natur.435..629S}. Each galaxy in the \citet{2006MNRAS.366..499D} catalog 
is characterized by a list of parameters for the bulge and the disk component of the galaxy separately
\footnote{The parameters include redshift, color (B, V, R, I, and K-band magnitudes), size (disk size 
estimates), dust estimates, and stellar population age estimates.}. A sophisticated fitting program 
then finds the best fit Spectral Energy Distribution (SED) parameters and galaxy extinction that reproduce 
the color information for each galaxy.
\catsim generates galaxy catalogs with realistic galaxy morphologies\footnote{The galaxies are modeled by 
Sersic profiles \citep{1968adga.book.....S,2013MNRAS.430..330H}, where the bulge and disk components have 
Sersic index 4 and 1, respectively.}, 
apparent colors and spatial distributions, and redshifts extending up to 5 on an area of $4.5 \times 4.5$ square degrees. 
The simulations include galaxies with $r$-band AB magnitudes brighter than 28. In \citet{A13}, the galaxy 
number density as a function of magnitude and redshift in the \catsim catalog is shown to be well matched to 
observations in the DEEP2 Redshift Survey\footnote{\url{http://deep.ps.uci.edu/}} \citep{2003SPIE.4834..161D, 
2004ApJ...617..765C}. The magnitude and size of each galaxy will be used to calculate the signal-to-noise ratio (\Eref{eq:snr}) 
and effective size (\Eref{eq:effR}) of each measurement (see \Aref{sec:derivation}). These two quantities determine the 
measurement noise, $\sigma_{m}$, for each galaxy (see \Sref{sec:sigmaM}). 

\begin{figure*}
 \begin{center}
 \subfigure[]{\includegraphics[scale=0.4]{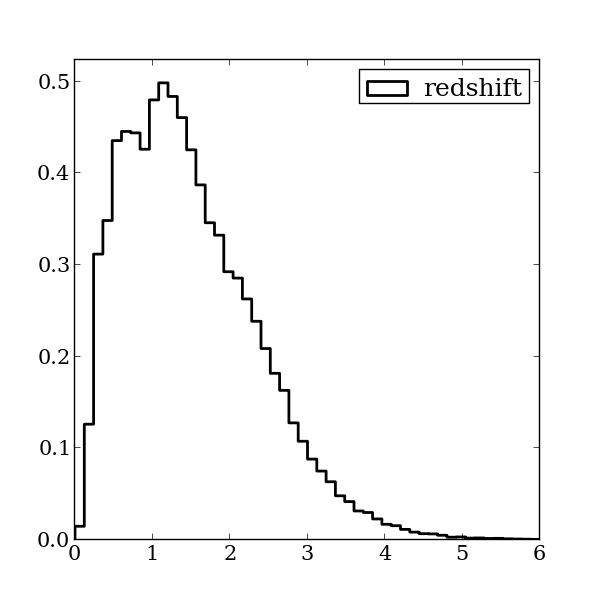} }
 \subfigure[]{\includegraphics[scale=0.4]{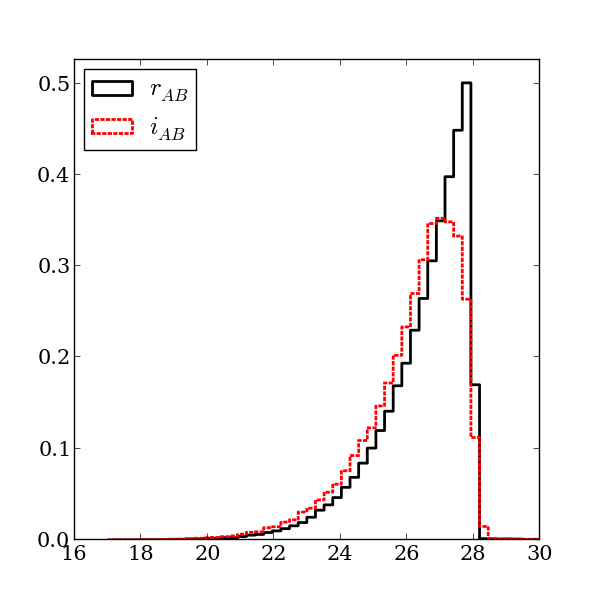}} \\
  \subfigure[]{ \includegraphics[scale=0.4]{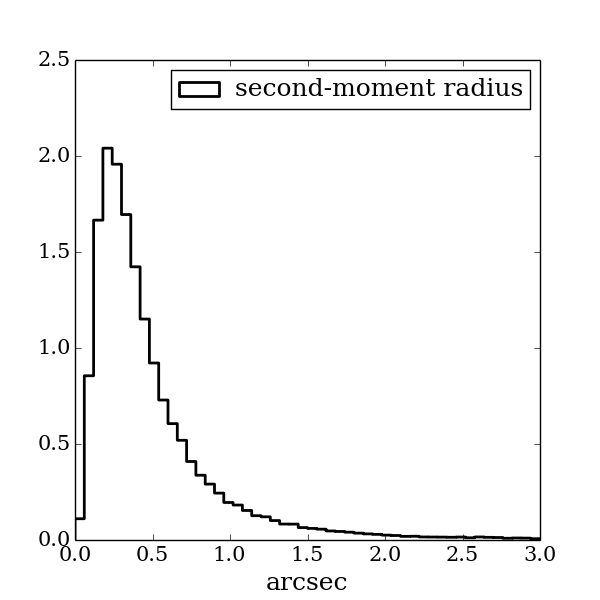}}
    \subfigure[]{ \includegraphics[scale=0.4]{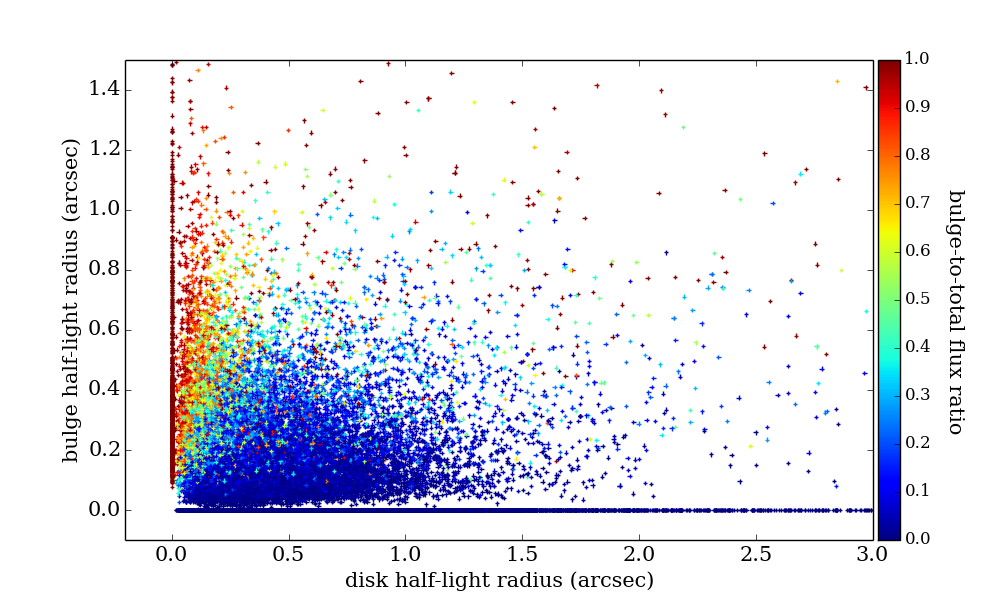}} \\ 
  \subfigure[]{\includegraphics[scale=0.4]{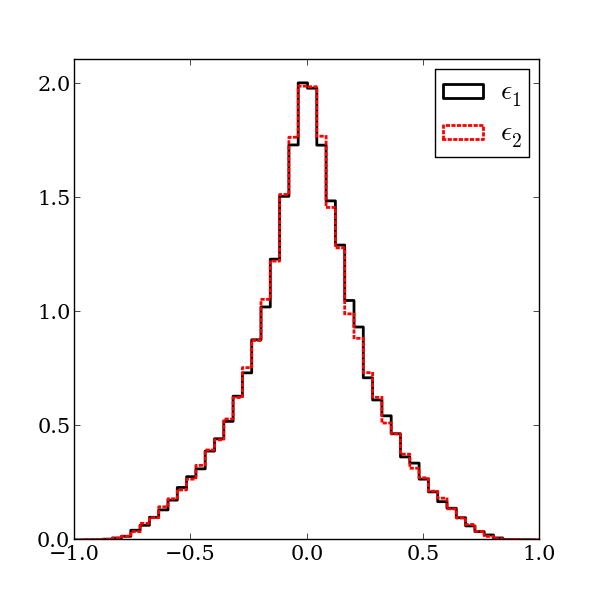} }
  \subfigure[]{\includegraphics[scale=0.4]{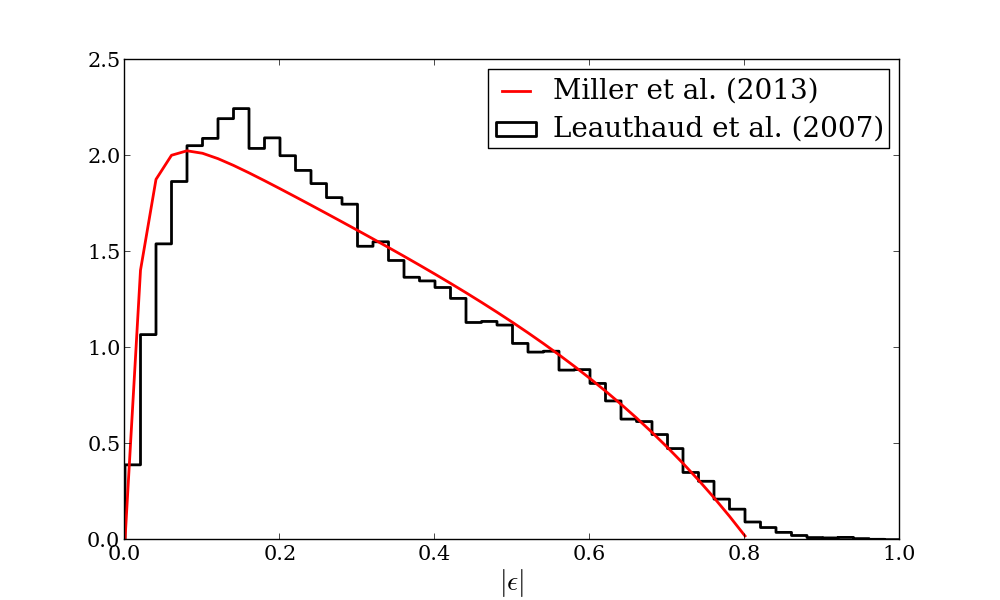} }\\ 
  \end{center}
 \caption{Distributions of (a) redshift, (b) $r$- and $i$-band magnitude, (c) size, and (e) single-component ellipticity 
in the galaxy catalog used for this study. All histograms are normalized to unit area. The sizes of the galaxies are 
measured by the second-moment radius, which can be derived from the half-light radius of the bulge and disk component and 
the bulge-to-disk flux ratio (\Aref{sec:galaxy_param}). In (d) we show how the half-light radius of the bulge and disk 
component and the bulge-to-disk flux ratio are distributed in the galaxy catalog. Notice that there are bulge-only and disk-only 
galaxies, which lie on the two axes in the plot. In (f) we compare the distribution of the absolute 
ellipticity with that derived in \citet{2013MNRAS.429.2858M}. } 
 \label{fig:catsim}
\end{figure*}

Finally, we assign shapes, or apparent ellipticities to each galaxy according to those measured in 
the COSMOS dataset\footnote{Since we are extracting the distribution from the COSMOS measurements directly, 
the ellipticity we assign to the galaxies include the galaxy intrinsic shape \textit{and} cosmic shear. However, we note 
that the level of cosmic shear is over an order of magnitude smaller than the level of the galaxies' intrinsic shape noise; 
\ie, \frefa{fig:catsim}(e) and (f) are dominated by shape noise.} \citep[][private communication]{2007ApJS..172..219L}.
These galaxy shapes have been corrected for measurement noise by excluding small and faint galaxies 
\citep{2007ApJS..172..219L}. In \Fref{fig:catsim}, we show the redshift, apparent $r$- and $i$-band (the two main lensing 
bands) AB magnitude, size and ellipticity ($\epsilon_{1}$, $\epsilon_{2}$) distributions of the galaxy population used in 
this study. The RMS width of \Fref{fig:catsim}(e) gives $\sigma_{SN}\approx 0.26$ for \Eref{eq:neff}. Finally, we note that the 
distribution of the absolute ellipticity ($|\epsilon|=\sqrt{\epsilon_{1}^{2}+\epsilon_{2}^{2}}$), as shown in 
\Fref{fig:catsim}(f), agrees well with that derived in \citet{2013MNRAS.429.2858M}, which is based on the SDSS dataset.

Although we assume shape noise to be independent of galaxy morphology, redshift, size and magnitude in our 
calculations, this is not strictly true. \citet{2009ApJ...700..791H}, for example, showed that the ratio of 
early-type (elliptical) to late-type (spiral and irregular) galaxies at low redshift is higher compared to that at high redshift. This
 could in principle introduce a redshift-dependent shape noise. However, as shown in 
\citet{2007ApJS..172..219L}, \citet{2013MNRAS.431..477J} and \citet{2013MNRAS.tmp.1297H}, the estimated shape noise as 
a function of galaxy redshift, size and magnitude is consistent with being flat within measurement noise in the COSMOS data. 
As a result, we choose to make the first-order approximation in this work that shape noise is constant as measured in 
\citet{2007ApJS..172..219L}. A non-constant $\sigma_{SN}$ correction to this approximation will require further investigation 
with deep space-based data \citep[\eg, ][]{2011PASP..123..596J} and better shear measurement algorithms.

\begin{figure*}
 \begin{center}
 \subfigure[]{\includegraphics[scale=0.4]{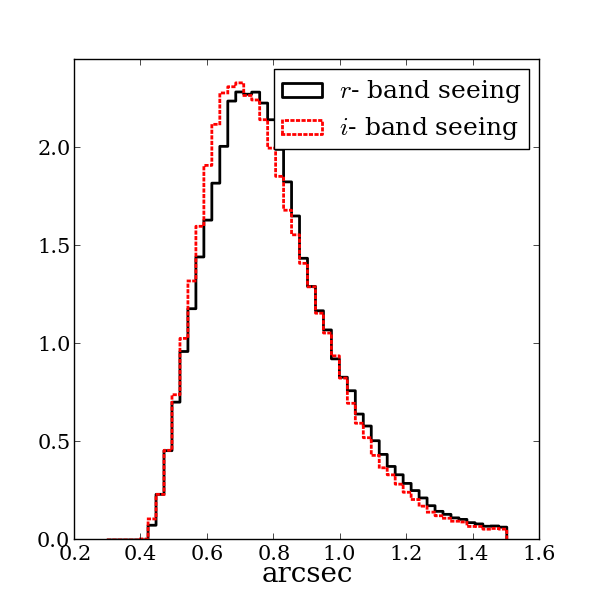} }
 \subfigure[]{\includegraphics[scale=0.4]{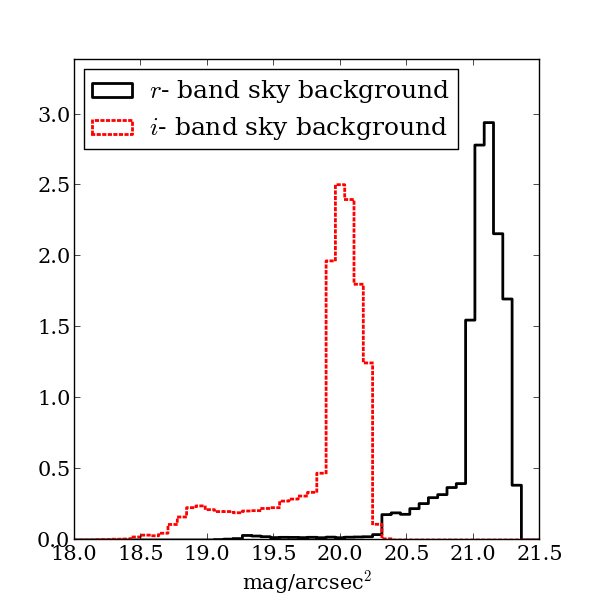}}
 \end{center}
 \caption{Distributions of (a) $r$- and $i$-band atmospheric seeing and (b) sky background for the expected 10-year LSST 
 survey as simulated from \opsim. All histograms are normalized to unit area.} 
 \label{fig:opsim}
\end{figure*}

\section{The shear measurement noise}
\label{sec:sigmaM}

In this section we estimate the shear measurement noise, $\sigma_{m}$, for a large range of galaxies under 
realistic observing conditions using high-fidelity simulations. We then select the galaxies used for weak lensing 
measurements. 

\subsection{Simulation and shear measurement}
\label{sec:phosims}
We invoke the LSST Photon Simulator v3.2\footnote{\url{http://dev.lsstcorp.org/trac/wiki/IS_phosim}} 
\citep[][\phosim]{P13} to generate high-fidelity images with realistic noise and instrumental/atmospheric effects. 
We also use the LSST Operations Simulator\footnote{\url{http://ssg.astro.washington.edu/elsst/opsim.shtml}} 
\citep[][\opsim]{2006SPIE.6270E..45D} catalog v3.61 to derive the expected distribution of observing conditions. 
\phosim is a fast Monte Carlo photon ray-tracing code that simulates all the major physical effects from the atmosphere 
down to the CCD readout. It adopts an atmospheric model with 7 Kolmogorov turbulent screens distributed from 
a few meters to a few tens of kilometers above the telescope. Each screen is described by several parameters associated 
with the wind speed/direction and turbulence strength. The optics model in \phosim is based on the most up-to-date 
engineering design, with optical errors (\eg, optics element mis-alignment, mirror surface perturbation, tracking errors, 
\etc) at the level set by engineering specifications. \opsim, on the other hand, models the telescope configuration, 
slewing mechanism, weather and sky conditions at the LSST site in a 10-year period. \Fref{fig:opsim} shows the 
atmospheric seeing\footnote{On top of the atmospheric seeing, we add the instrumental PSF 
$\approx$ 0.4" in quadrature to get the total PSF size used in later calculations.} and sky background distribution in $r$- and 
$i$-band from the \opsim catalog used in this paper. Due to the wavelength-dependent nature of the system throughput and 
PSF size, the best possible image quality (maximum signal-to-noise ratio + minimum PSF size) is usually 
achieved in the $r$- and $i$-band images. For optimal results in weak lensing measurements, the \opsim algorithm would thus 
preferentially choose to image in one of these ``lensing bands'' when the observing conditions are good. 

\begin{figure*}
 \begin{center}
\subfigure[]{\includegraphics[scale=0.5]{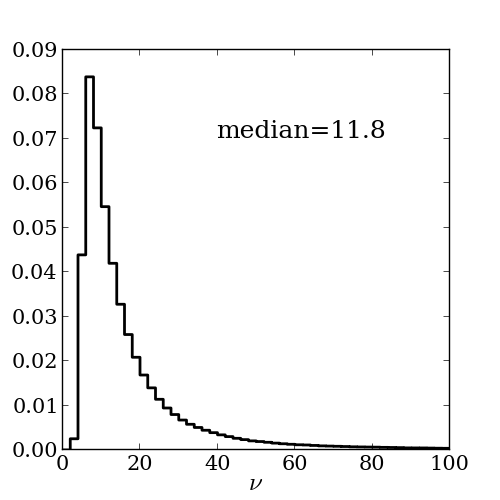}}
\subfigure[]{\includegraphics[scale=0.5]{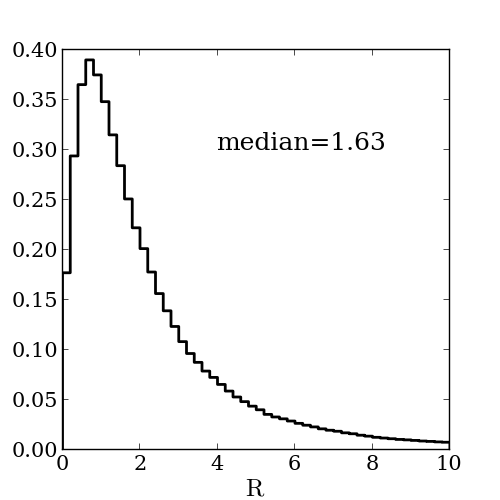}}
 \end{center}
 \caption{Distribution of the (a) signal-to-noise ratio ($\nu$) and (b) effective size ($\rm R=r_{\rm gal}^2/r_{\rm PSF}^2$) 
 for galaxies that we use to estimate the shear measurement uncertainty $\sigma_{m}$. These are sources detected in 
 a simulated single 15s exposure with an $r$-band filter, and have well-defined shear estimates (i.e. this is \textit{not} 
 the $\nu$ and $R$ distribution for all sources detected in a single exposure).  
 The median values of these distributions are listed on the plots and all histograms are normalized to unit area.} 
 \label{fig:nu_R_histo}
\end{figure*}

With the input galaxy catalog from \catsim and the observing parameters from \opsim, we use \phosim to generate a 
set of 1,000 simulated 15-s $r$-band LSST exposures\footnote{In the nominal survey plan for LSST, the telescope will take two 
15-s exposures separated by 2--3-s readout with all instrument configurations held fix. This full sequence is called a ``visit'', 
which is effectively a continuous 30-s exposure. To model this, we estimate $\sigma_{m}$ in single 15-s exposures from the 
simulations and assign identical observing parameters for the two exposure in the same visit.} 
on a single LSST CCD sensor. For the 1,000 exposures, we simulate galaxies with the redshift, size, magnitude and shape 
distributions\footnote{We sample galaxy parameters from the 3D redshift-size-magnitude distribution, and assign a random ellipticity from 
\Fref{fig:catsim}(d).} in \Fref{fig:catsim}, and randomly assign observing conditions based on the distributions in \Fref{fig:opsim}. We select 
10 random locations on the LSST focal plane for these simulations to capture realistic PSF effects across the field of view. For each 
exposure, we make two images: The first image contains galaxies from \catsim, while the second image is identical to the first, except 
that we replace each of the galaxies with a bright star that effectively samples the PSF of its galaxy counterpart in the 
first image.

For each simulated image, we detect objects using the Source Extractor software \citep{1996A&AS..117..393B} and 
measure shear using the \imcat software package\footnote{\url{http://www.ifa.hawaii.edu/~kaiser/imcat/}. The shear 
measurement algorithm in \imcat is based on the KSB algorithm developed by \citet{1995ApJ...449..460K, 1997ApJ...475...20L} 
and \citet{1998ApJ...504..636H}.}. The shear measurement algorithm \imcat performs the following steps: First, the shapes of 
both stars and galaxies are measured. Then, for each galaxy in the first image, the PSF effects are corrected using the 
corresponding bright star in the second image. Finally, a shear value is calculated for each galaxy. The difference between the measured 
shear and the input shear is the shear measurement error, or 
\begin{linenomath*}\begin{equation} 
\delta \boldsymbol{\gamma} = \boldsymbol{\gamma_{\rm measured}} - \boldsymbol{\gamma_{\rm input}}\;\;.
\label{eq:del_g}
\end{equation} \end{linenomath*}
We use \imcat for the shape/shear measurement because it is one of the most commonly used methods for weak lensing studies. 
The results from this paper will thus be directly relevant to the ongoing and future surveys. But given recent improvements in the 
shear measurement methods \citep{2009AnApS...3....6B, 2012MNRAS.423.3163K, 2012arXiv1204.4096K, 2013arXiv1302.0183Z, 
2013MNRAS.429.2858M, 2013arXiv1304.1843B}, our results are conservative in this regard. 

Note that in this  analysis framework we have ignored several effects from a realistic stellar population. First, we choose to 
estimate $\neff$ for a typical weak lensing field, with Galactic latitude of $|b|\sim60$ degrees. The stellar density in these fields is 
relatively low ($\approx 1$ arcmin$^{-2}$), such that significant blending from stars can be ignored. Second, by simulating the 
``true'' PSF for each galaxy via simulations (in the second image), we have avoided the process of PSF modeling. This includes 
selecting well-measured stars and modeling the PSF by interpolating shape parameters of these stars. The effect of imperfect star 
selection is expected to be small, since the bright stars used for PSF modeling are generally easy to identify without much 
contamination from the galaxies. The effect of interpolating shape parameters from stars depends on the interpolation scheme. In 
\Aref{sec:truePSF}, we show that using a conventional interpolation method for PSF modeling yields only a small (3--6\%) 
degradation in $\neff$ compared to the case for which the PSF is known exactly. 

Our approach of estimating shear measurement noise from \phosim is very similar to that used in \citet{2013arXiv1301.0830B}. 
The major differences between the simulations described here and that in \citet{2013arXiv1301.0830B} are (1) we simulate 
single-exposure depths instead of the expected full survey depth, and (2) we do not include shape noise in our definition of 
measurement errors.

\subsection{Single exposure shear measurement noise for a single galaxy}
\label{sec:single}

To characterize the shear measurement noise $\sigma_{m}$, we follow the procedure suggested by \citet{2002AJ....123..583B}, and assume 
that shear measurement noise ($\sigma_{m}$) is mainly determined by the signal-to-noise ratio ($\nu$) and the effective size ($\rm R$) 
of the measured galaxy via the following form: 
\begin{linenomath*}\begin{equation} 
\sigma_{m}(\nu, {\rm R})=\frac{a}{\nu} \left[1+\left( \frac{b}{{\rm R}} \right)^{c}\right]\; ,
\label{eq:fit_sigmam}
\end{equation} \end{linenomath*}
\noindent where ($a,b,c$) are coefficients depending on the shear algorithm and the image quality. 

Here, the galaxy's signal-to-noise ratio ($\nu$) is defined as
\begin{linenomath*}\begin{equation} 
  \nu=\frac{S}{\sqrt{S+B}}\; ,
\label{eq:snr}
\end{equation} \end{linenomath*}
where $S$ is the source photon counts and $B$ is the background photon counts\footnote{In practice, both of these are calculated 
in an aperture -- we use the convention adopted by Source Extractor, where they define the aperture size $r_{ap}$ to be twice 
the first-moment radius of the total light distribution \citep[\Eref{eq:r_ap},][]{1980ApJS...43..305K}. This aperture size has been shown 
to contain $>90\%$ the total signal, for a wide range of galaxy shapes.}. 
And the galaxy's effective size ($\rm R$) is defined to be the relative size of the galaxy to the PSF, or  
\begin{linenomath*}\begin{equation} 
{\rm R}=\frac{r_{\rm gal}^2}{r_{\rm PSF}^2}\;.
\label{eq:effR}
\end{equation} \end{linenomath*}
Here, $r_{\rm gal}$ and $r_{\rm PSF}$ are the second-moment radii (see \Eref{eq:2nd_moment}) of the galaxy and the PSF's light 
distribution. 
In \Aref{sec:derivation}, we show how $\nu$ and $\rm R$ can be calculated from observational quantities as well 
as the generic galaxy model described in \Sref{sec:sigmaSN} and observational parameters.

\begin{figure}
 \begin{center}
\includegraphics[height=2.8in]{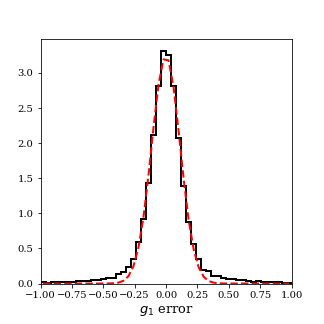}
\end{center}
\caption{Distribution of the shear measurement error in the $\nu\sim15$, $\rm R\sim1$ bin. The overlaid red dashed curve 
shows the best-fit Gaussian with a standard deviation of 0.11, which does not capture the wings of the distribution. We take 
the (unclipped) standard deviation of the histogram as an estimate for $\sigma_{m}$ for this bin, which is 0.32 in this example.
The shape of this distribution is typical for most ($\nu, \rm R$) bins.} 
 \label{fig:sigmam_histo}
\end{figure}
\begin{figure}
 \begin{center}
\includegraphics[height=2.8in]{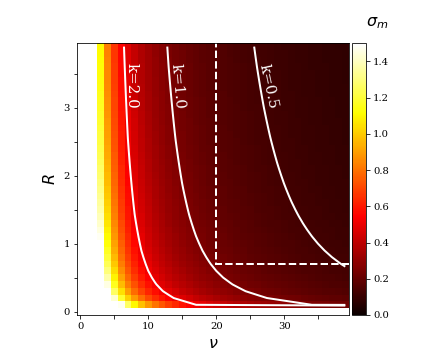}
 \end{center}
 \caption{Shear measurement noise ($\sigma_{m}$) as a function of galaxy signal-to-noise ratio ($\nu$) and effective size ($\rm R$). 
 As discussed in \Sref{sec:gal_selection}, the three solid contours correspond to the galaxy selection cuts (see \Eref{eq:k_def} 
 for definition of $k$) we apply. The dashed lines indicate a conventional 2D cut ($\nu>20$, $\rm R>0.7$) that approximately 
 shares the same maximum measurement error as the $k=1$ cut.} 
 \label{fig:measurement_noise}
\end{figure}

The main task of this section is to determine the coefficients in \Eref{eq:fit_sigmam} from the shear measurements in 
\Sref{sec:phosims}. To do this, we first calculate the $\nu$ and $\rm R$ values for all galaxies measured in \Sref{sec:phosims} and 
bin them into ($\nu$, $\rm R$) bins. Then, for all galaxies in each ($\nu$, $\rm R$) bin, we calculate the RMS of the shear measurement 
errors $\delta \boldsymbol{\gamma}$ (\Eref{eq:del_g}). This provides an estimate for $\sigma_{m}$ for that bin. We then fit the 
$\sigma_{m}(\nu, {\rm R})$ surface and determine the $(a,b,c)$ coefficients.

We use 30 logarithmic bins in $\nu$ ranging from 3 to 650 and 30 logarithmic bins in $\rm R$ ranging from 0.2 to 43, which covers 
all the galaxies measured in these simulations. Since each bin contains a different number of galaxies, an uncertainty estimate of 
$\sigma_{m}/\sqrt{N_{\rm bin}}$ is used for each bin during the fit, where $N_{\rm bin}$ is the number of galaxies in that bin.
The distributions of the $\nu$ (for any $\rm R$) and of $\rm R$ (for any $\nu$) measured from single-exposure simulations are shown 
in \Fref{fig:nu_R_histo}, and a typical histogram for one of the bins is shown in \Fref{fig:sigmam_histo}. As can be seen, the shear 
measurement error is slightly non-Gaussian with some low-level wings. 

We derive the best-fit coefficients of \Eref{eq:fit_sigmam} from the above procedure to be $(a,b,c)=(1.58, 5.03, 0.39)$. The RMS 
difference between the fit and the measured values (weighted by $N_{\rm bin}$) is $\sim0.1$. The fitted surface is shown in 
\Fref{fig:measurement_noise}. In general, the behavior of the measurement noise is intuitive -- small and faint galaxies have noisier 
measurements and large, bright galaxies are well measured. 

We can now estimate the shear measurement noise ($\sigma_{m}$) for each galaxy in each exposure by plugging the 
$\nu$ and $\rm R$ values (for each galaxy in each exposure) into \Eref{eq:fit_sigmam}. In this calculation, four parameters from 
the \catsim catalog (the apparent magnitude, the galaxy bulge and disk half-light radii, and the bulge-to-total flux ratio) and two 
parameters from the \opsim catalog (the seeing and sky background in each exposure) are used. Note that we have derived the 
model of shear measurement noise ($\sigma_{m}$) via galaxies in a certain $\nu$ and $\rm R$ range that can be detected and 
measured in single-exposure simulations (\Fref{fig:nu_R_histo}). For galaxy measurements with $\nu$ and $\rm R$ values 
outside this range, our estimation of $\sigma_{m}$ would be an extrapolation. 

\subsection{Combining multiple shear measurements of a single galaxy}
\label{sec:combine}

Shear measurements from the same galaxy imaged in multiple exposures are combined to give a single estimate of 
shear per galaxy. For LSST, this is a crucial step because the number of exposures of each galaxy is typically an order of 
magnitude larger than for previous surveys. That is, the algorithm one uses to combine these shear measurements is important. 

Different approaches have been suggested to deal with such multi-epoch datasets. Conventionally, one would create 
co-added images and carry out the full analysis on the co-add. This is not an optimal approach, for one throws away information 
obtained in the sharpest images and furthermore, correlates the noise in the pixels and creates biases in the shear estimates. 
\citet{2002AJ....123..583B} and 
more recently \citet{2007MNRAS.382..315M} and \citet{2008ASPC..394..107T} have suggested taking the approach of joint fitting, 
where the individual exposures are kept separate throughout the analysis. The detections would still be made on a coadded image to 
maximize signal-to-noise ratio in the detection process.  But then the shape measurement would involve a joint fit using all the original pixels from 
the individual exposures, making it possible to weight the measurement in different exposures according to the noise on that particular 
exposure. That is, we extract more information from good images and less information from bad images. The latter approach is 
optimal when the image quality varies across exposures.

First, we consider the case for which the shear is estimated jointly using all the original (individual) images.  
In this case, the optimal joint estimator will have a net measurement error, $\sigma_{m,{\rm joint}}$, of
\begin{linenomath*}\begin{equation} 
\sigma_{m,{\rm joint}}=\left[ \Sigma_{j}^{N_{\rm exp}} \frac{1}{\sigma_{m,j}^{2}} \right]^{-0.5}\;\;,
\label{eq:jf_sum}
\end{equation} \end{linenomath*}
where $\sigma_{m,j}$ is the measurement error estimate from \Eref{eq:fit_sigmam} for each exposure $j$ out of the 
$N_{\rm exp}$ total exposures.

For the second case of combining the images and then measuring shear from the coadded image, 
we can calculate the effective $\nu$ and $\rm R$ values using the total signal and background and 
an estimate of the net PSF size from adding the $N_{\rm exp}$ exposures:
\begin{linenomath*}\begin{align} 
\label{eq:coadd}
&S_{\rm tot} = \Sigma_{j}^{N_{\rm exp}} S = N_{\rm exp} S\; , \\ \notag
&B_{\rm tot} = \Sigma_{j}^{N_{\rm exp}} B_{j} \; ,   \\ \notag
&r^2_{\rm PSF, eff}=\frac{1}{N_{\rm exp}} \Sigma_{j}^{N_{\rm exp}} r^2_{{\rm PSF},j}\; , \\ \notag
& \nu_{\rm eff}=\frac{S_{\rm tot}}{\sqrt{S_{\rm tot}+B_{\rm tot}}}\; , \\ \notag
&{\rm R_{eff}}=\frac{r_{\rm gal}^2}{r_{\rm PSF, eff}^2}\;,
\end{align} \end{linenomath*}
and
\begin{linenomath*}\begin{equation} 
\sigma_{m,{\rm coadd}}= \frac{a'}{\nu_{\rm eff}} \left[1+ \left( \frac{b'}{{\rm R_{eff}}} \right) ^{c'} \right] \; .
\label{eq:sigmam_coadd}
\end{equation} \end{linenomath*}
In principle, $(a',b',c')$ and $(a,b,c)$ need not be identical. The difference can be due to the specific PSF modeling technique 
one uses.
This means that the fit shown in \Fref{fig:measurement_noise} (based on single exposure measurements) may no 
longer be appropriate in the case of coadded images. Nevertheless, for the purpose of this analysis, we avoided most PSF-related issues 
and specifically explored a wide range of reasonable $(\nu, {\rm R})$ values. As a result, we will make the approximation 
$(a',b',c')\approx(a,b,c)$, effectively generalizing \Eref{eq:fit_sigmam} to all $(\nu, {\rm R})$ values. 

We first consider the case for which only $r$-band images are used and then consider the case for which both $r$- and $i$-band images 
are used. For LSST, we expect a similar number ($\approx$ 368) of $r$- and $i$-band 15-s exposures in the full 10-year dataset. 
Note that the technical difficulties of combining images for different bands still need to be assessed (see also discussion in \Sref{sec:multi_filter}).

\subsection{Galaxy selection}
\label{sec:gal_selection}

In all weak lensing analyses to date, one does not use all the galaxies that are detected and identified as galaxies to conduct 
cosmic shear measurements. Instead, only galaxies that pass certain criteria make it into the final summation in 
\Eref{eq:neff}. Although noisy galaxies naturally have low weights in \Eref{eq:neff}, there are several practical reasons that 
one rejects part of the galaxy population: First, most shear measurement algorithms become numerically unstable 
when working with noisy galaxies. Second, these noisy galaxies are subject to noise bias 
\citep{2012MNRAS.425.1951R, 2012MNRAS.424.2757M} and can increase the systematic errors significantly. Finally, 
assuming one were to calibrate these systematic errors with simulations \citep{2012MNRAS.427.2711K}, the calibration 
process becomes challenging and unstable as the measurement noise increases. 

Given the reasoning above, a natural route to select galaxies is to base the selection on the relative level of the shear 
measurement errors ($\sigma_{m}$) and the shape noise ($\sigma_{SN}$). In other words, we use only galaxies that satisfy:
\begin{linenomath*}\begin{equation} 
\sigma_{m} < k \:\sigma_{SN},
\label{eq:k_def}
\end{equation} \end{linenomath*}
where $k$ is of order unity and depends on the performance of the shear measurement algorithm. In the main analysis of this paper, 
we choose three $k$ values based on studies in \citet{2010MNRAS.405.2044B} and \citet{2012MNRAS.423.3163K}. These 
studies have shown that most current shear measurement algorithms perform well on objects with 
($\nu,\rm R$)$\approx$ (40, 1.5), operate with moderate accuracy on objects with ($\nu, \rm R$) $\approx$ (20, 1), and tend to fail 
on objects with ($\nu,\rm R$) $\approx$ (10, 0.5). These three cases roughly correspond to $k=$(2.0, 1.0, 0.5) if we assume the 
shear measurement noise follows \Fref{fig:measurement_noise}. In the rest of this paper, we will thus consider these three 
galaxy selection cuts, where $k=1.0$ corresponds to the fiducial case, $k=2.0$ corresponds to the optimistic case, and $k=0.5$ 
corresponds to the conservative case. The fiducial case corresponds to using a shear measurement algorithm with accuracy 
similar to current state-of-the-art methods. 

We note that a more common approach in existing weak lensing analysis pipelines is to select galaxies 
based on a 2D cut in the $\nu$-$\rm R$ plane. However, we argue that galaxy selection based on such a cut is not necessarily 
optimal. As illustrated in \Fref{fig:measurement_noise}, the two galaxy samples selected by the $k=1$ contour and the 
dashed rectangle (equivalent to a 2D cut of $\nu>20$, ${\rm R}>0.7$) have similar allowed maximum measurement noise. But the 
dashed line clearly removes less parameter space than the $k=1$ contour, thus leading to smaller $\neff$. This simple illustration 
demonstrates that if the measurement noise can be properly estimated, the most efficient way to select galaxies for weak lensing 
measurements is to base the selection on the measurement noise directly (\Eref{eq:k_def}). 

Finally, we consider the range of redshifts for galaxies used in the analysis. It is common for lensing analyses to only consider a limited redshift 
range since objects at very high redshift will have photometric redshifts that are poorly estimated, while objects at very low redshift do 
not provide much cosmological information. We assume that the LSST photometric redshifts uncertainties are consistent with those given 
in \citet{2009arXiv0912.0201L}\footnote{The question of whether near infrared observations will be required to achieve this level of accuracy 
or if advanced methods such as those proposed by \citet{2008ApJ...684...88N} will be sufficient is beyond the scope of this paper.} and consider 
only galaxies in the redshift range 0.1-- 3. 

\begin{figure}
 \begin{center}
  \subfigure[]{\includegraphics[height=2.5in]{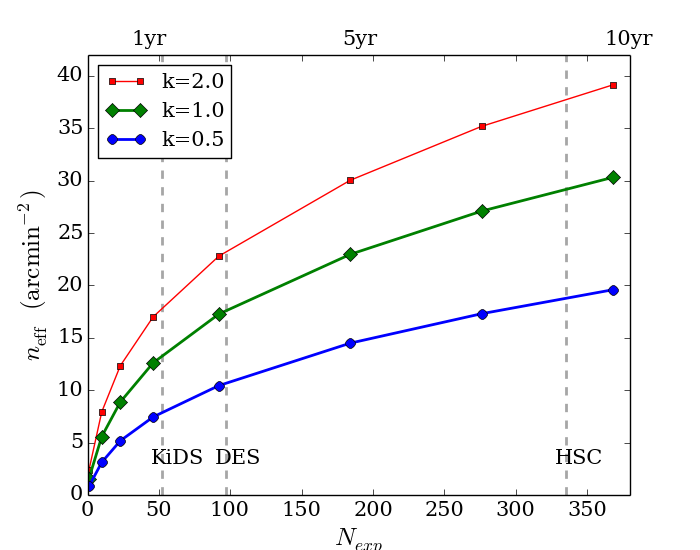} }
  \subfigure[]{\includegraphics[height=2.5in]{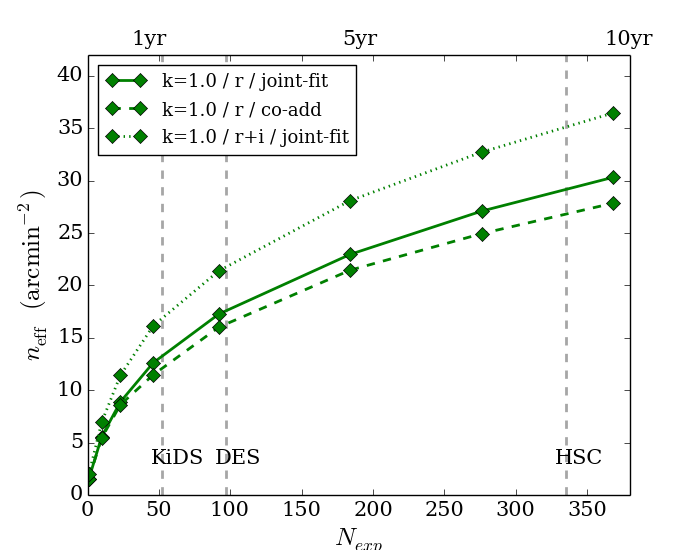}}
 \end{center}
 \caption{$\neff$ as a function of the number of exposures, or operation time (as listed on the top axis). In (a), we 
 show the case for which $r$-band data are combined with a joint-fitting approach and with different galaxy selection 
 cuts (\Eref{eq:k_def}). The red, green and blue curves are for the optimistic ($k=2.0$), 
 fiducial ($k=1.0$), and conservative ($k=0.5$) scenario, respectively. In (b), we show for the fiducial case, how the 
 curves change when one combines multiple exposures via a co-add method (dashed curves) and when multi-band 
 data are included (dotted curves). The three vertical dashed lines show the approximate $N_{\rm exp}$ values 
 corresponding to the equivalent depth of the three ongoing surveys: KiDS, DES and HSC.} 
 \label{fig:neff_N_all}
\end{figure}

\section{Estimation of $\neff$ for LSST}
\label{sec:neff}

We now estimate $\neff$ by combining the analyses from \srefa{sec:sigmaSN} and \srefb{sec:sigmaM} into \Eref{eq:neff}: First, 
we estimate the shear measurement error on each galaxy in the galaxy catalog after combining $N_{\rm exp}$ exposures. Then we 
reject galaxies with measurement errors and redshift values that fail the selection criteria set in \Sref{sec:gal_selection}. 
The remaining galaxies are used to calculate $\neff$ using \Eref{eq:neff} and $\sigma_{SN}=0.26$. We consider two combining 
approaches (co-add and joint-fit), and also the possibility of combining $r$- and $i$-band data. \Tref{tab:cut_neff} and 
\Fref{fig:neff_N_all} summarize the results of our analysis. We discuss below several issues related to estimating $\neff$.

\begin{table}
  \centering
  \caption{Summary of the effective number of galaxies per square arcminute used for weak lensing analyses, or $\neff$, 
  derived in \Sref{sec:neff} for the 10-year data from LSST. The columns under $n$ list the raw galaxy number densities for 
  different galaxy selection cuts (\Eref{eq:k_def}), while the columns under $\neff$ list the corresponding effective galaxy 
  number densities. In the 
  first row, we use $r$-band images only and combine the multiple exposures via joint-fitting. The second row shows how 
  the numbers change as we adopt a co-add approach instead. The last row shows the case for which both $r$- and $i$-band 
  images are used. The bold face values are used later in \Sref{sec:practicalities}.}
   \begin{tabular}{ lcc cccc }
     & \multicolumn{3}{c}{$n$} &  \multicolumn{3}{c}{$\neff$} \\ \hline
     $k$  &2.0 & 1.0 & 0.5                      &2.0 & 1.0 & 0.5  \\ \hline
     $r$ / joint-fit         & 64   & 38   & 21  & 39              & 30              & 20  \\ 
     $r$ / co-add         & 59  & 35    & 19  & 36               & 28              & 18  \\ 
     $r$+$i$ / joint-fit  & 78   & 46     & 26  & \textbf{48} & \textbf{37} & \textbf{24}\\    
  \end{tabular}
  \label{tab:cut_neff}  
\end{table}

\subsection{Time dependence of $\neff$}
In \Fref{fig:neff_N_all}, we show $\neff$ as a function of number of exposures combined, or, equivalently, survey time. 
\Fref{fig:neff_N_all}(a) shows the $\neff$ behavior for $r$-band only and for different $k$ values as described in 
\Sref{sec:gal_selection}. \Fref{fig:neff_N_all}(b) shows, for the fiducial case ($k=1.0$), how the result will change as one 
considers combining multiple exposures differently and when $i-$band data are included. The general trends for all plots are 
similar: $\neff$ increases monotonically and does so faster in the beginning of the survey. In all cases studied here, the 
curves do not plateau during the survey. 

As the exposure time increases, the same object will be measured with larger signal-to-noise ratio on the combined image and is more 
likely to survive the $\sigma_{m}$ cut. Given that the number of galaxies increases dramatically as one goes to the fainter 
end of the galaxy population (see \Fref{fig:catsim}(c)), we can expect a sharp increase in $\neff$ over time. 
However, as the number density of galaxies surviving the $\sigma_{m}$ cut increases, blending may become an issue. 
We estimate in \Sref{sec:practicalities} the possible effect of blending on $\neff$. We also note that the rate of increase of $\neff$ 
is slower than the naive expectation of $\sqrt{N_{\rm exp}}$. This is because we place a cut on $\sigma_{m}$ and not simply 
on signal-to-noise ratio. Thus $\neff$ will not necessarily scale with $\sqrt{N_{\rm exp}}$.

In \Fref{fig:neff_N_all}(b), we first look at the effect of combining multiple exposures using a more traditional co-add method 
versus the joint-fitting method described in \Sref{sec:combine} (comparing the solid curves and the dashed curves). The two 
curves start off at similar levels. Approximately half a year into the survey, $\neff$ becomes slightly larger for the joint-fit approach. 
This is due to the fact that the joint-fit method is optimal at extracting information from multiple exposures with very different 
observing conditions, and that effect becomes more pronounced as one collects more data. 
When including the $i$-band data, a fractional increase of 20--30\% in $\neff$ is shown 
throughout the survey period (comparing the solid curve and the dotted curve).  

To compare the performance of LSST with ongoing surveys, we also label on the plot the ``equivalent $N_{\rm exp}$ values'' 
for KiDS, HSC and DES. These are rough estimations based on the $r$-band limiting magnitudes for the three surveys. We expect 
the source flux corresponding to the limiting magnitude to scale inversely with the square-root of the survey time, and thus $N_{\rm exp}$. 
In other words,
\begin{linenomath*}\begin{equation} 
10^{-0.4({\rm m^{X}}-{\rm m^{LSST}})} = \sqrt{\frac{N^{\rm LSST}_{\rm exp}}{N^{\rm X}_{\rm exp}}}, 
\label{eq:N_scal}
\end{equation} \end{linenomath*}
where $\rm m$ is the $r$-band limiting magnitude and the superscript denotes the survey of interest. 
For LSST, $\rm m^{LSST}=27.5$ and $N^{\rm LSST}_{\rm exp}=368$. Given the limiting magnitudes for the three surveys: 
25.2 (KiDS), 27 (HSC) and 25.6 (DES), \Eref{eq:N_scal} yields: $N^{\rm KiDS}_{\rm exp}=52$, $N^{\rm HSC}_{\rm exp}=335$ 
and $N^{\rm DES}_{\rm exp}=97$.
Note that this estimation does not account for the different image quality and measurement errors in the different surveys. 

\subsection{Effect of weighting and galaxy selection cut}

As expected from \Eref{eq:neff}, $\neff$ is always smaller than the raw number of galaxies surviving the $\sigma_{m}$ cut, or $n$. We 
compare the first row in the $n$ and $\neff$ columns of \Tref{tab:cut_neff}. We find that $\neff$ is 61\%, 79\%, 95\% of $n$ for 
$k=$2.0, 1.0 and 
0.5, respectively. This rapid increase in the $\neff$-to-$n$ ratio is sensible: Smaller $k$ cuts suggest a lower-noise galaxy sample, and 
that means each galaxy will contribute more to $\neff$. Low measurement noise is key to a larger $\neff$ value, for it affects $\neff$ in 
two ways: (1) it enables small and faint galaxies to pass the $\sigma_{m}$ cut and (2) it weights individual galaxies 
more in the summation in \Eref{eq:neff}. 

We also examine the effects of using a conventional two-dimensional galaxy selection cut in the $\nu$-$\rm R$ plane. We calculate 
$\neff$ in the case for which the two-dimensional cut ($\nu>20$, $\rm R>0.7$) is used instead of the measurement noise cut 
$\sigma_{m}<\sigma_{SN}$ ($k=1$). As shown in \Fref{fig:measurement_noise}, the two-dimensional cut corresponds to a smaller 
parameter space on the $\nu$-$\rm R$ plane. This reduction in parameter space results in a significant ($\sim25\%$) decrease in 
$\neff$. 

Finally, the redshift selection cut we apply reduces $\neff$ only slightly, at the $<$1\% level. 

\subsection{Co-add vs. joint-fit}
In practice, there are several incentives to combine multiple exposures via a joint-fit method rather than a co-add method. This 
includes avoiding the process of homogenizing the multiple exposures and reducing correlated noise in the images 
\citep{2013MNRAS.429.2858M}. The statistical power is not commonly considered when making this choice. However, using a 
joint-fit method naturally yields higher statistical power compared to using a co-add method. The improvement comes from the 
fact that for a co-add method, only galaxies well measured in the ``average'' exposure are used, while this is not necessarily true 
for the joint-fit method. It is possible to use galaxies that are not well measured in the ``average'' exposure in a joint-fit method, as 
long as the galaxies are sufficiently well measured in some of the exposures. We have estimated in \Tref{tab:cut_neff} a small 
decrease in $\neff$ ($\sim7\%$) going from a joint-fit to a co-add approach. We note that in this calculation, our $\neff$ estimation 
for the co-add approach is less accurate than that for the joint-fit approach. This is because, as noted in \Sref{sec:combine}, we 
have extrapolated \Fref{fig:measurement_noise} from single-exposure results to the co-add regime. In the co-add images, not only 
does the photon noise in the galaxies decrease, the stars also become better measured and the PSF model becomes smoother and 
easier to model. But since we have avoided PSF estimation in this analysis, \Fref{fig:measurement_noise} should be a good approximation 
for both the single exposure and the co-added image.

\subsection{Combining data from multiple bands}
\label{sec:multi_filter}
In most existing weak lensing analyses, only single-band data are used, as the surveys are usually designed to have the 
best image quality data in a single band. For LSST, both $i$- and $r$-band data should be sufficiently good 
for lensing analyses. We thus consider the case for which both $i$- and $r$-band images are used. Comparing the first and the last row in 
\Tref{tab:cut_neff}, we see that the addition of $i$-band data results in a $\sim$23\% growth in $\neff$ and not a naive $\sqrt{2}\approx 1.4$ 
factor increase \citep{2008JCAP...01..003J}. This is because $i$-band images generally have lower signal-to-noise ratio 
(higher background as seen in \Fref{fig:opsim}) and thus higher measurement noise. We also estimate $\neff$ for the case of combining all six 
filter bands using the same method. We find that $\neff \approx$ 68, 54 and 36 arcmin$^{-2}$ for the optimistic, 
fiducial and conservative 
cases, respectively. That is, $\neff$ increases by a factor of $\sim1.8$ going from one filter ($r$) to six filters and $\sim$1.5 going from 
two filters ($r$+$i$) to six filters. Again, the gain is smaller than $\sqrt{6}\approx 2.4$ and $\sqrt{3}\approx 1.7$. In addition to the fact that all 
other bands are shallower than $r$-band, the average seeing in other bands is generally similar or worse than the $r$- 
and $i$-band seeing. 

In the estimation of $\neff$ above for multi-band datasets, we have made several assumptions. First, we assumed 
that individual galaxy shapes are the same in the different filter bands. This is a good assumption according to 
\citet{2008JCAP...01..003J}, who showed that the shape measurements in deep multi-band space data are highly correlated between 
different filter bands. 
Second, we have assumed that the measurement noise in each band depends on the galaxy's signal-to-noise and effective size 
in the same way as the $r$-band images (\ie, the coefficients of \Eref{eq:fit_sigmam} are the same). This assumption could fail if, for 
example, the PSF modeling algorithm does not perform equally well in all bands. However, since we have minimized the effect 
from the PSF modeling procedure in our analysis, this should be a good assumption. Finally, we made the assumption that the 
joint-fitting algorithm is capable of operating on the multi-band dataset. 

Given the assumptions discussed above, a more detailed study is required to understand the actual quantitative gain in 
combining data from multiple filter bands. Nevertheless, we point out here that it is in principle possible to achieve large 
$\neff$ values by combining data from all the available filters, even if some filters are not optimized for lensing. This 
conclusion is consistent with that found in \citet{2008JCAP...01..003J}.

\subsection{Redshift distribution of $\neff$}

\begin{figure}
 \begin{center}
 \includegraphics[height=2.3in]{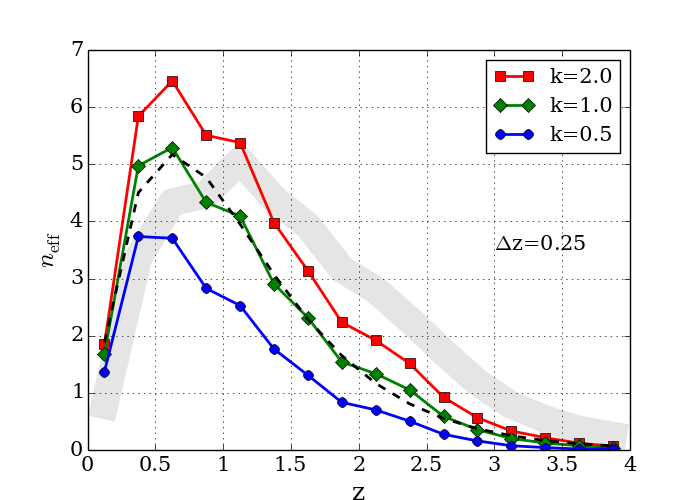} 
 \end{center}
 \caption{For the optimistic ($k=2.0$, red squares), fiducial ($k=1.0$, green diamonds) and the conservative ($k=0.5$, blue 
 circles) scenarios, $\neff$ calculated in different redshift bins before the $0.1<z<3$ redshift cut is applied. Each redshift bin spans 
 an interval of $\Delta z=$0.25. The grey band shows the raw galaxy number density $n$ before any galaxy selection cut is applied, 
 normalized to similar amplitudes as the other curves for qualitative comparison. The black dashed curve is the best-fit functional form 
 (\Eref{eq:z_dist}) for the fiducial case.} 
 \label{fig:neff_z}
\end{figure}

Finally, we look at the redshift distribution of $\neff$. In \Fref{fig:neff_z}, we plot $\neff$ in 16 redshift bins from 
redshift 0 to 4 for the optimistic, fiducial and conservative scenarios. Also plotted is the raw galaxy number 
density $n$ before any galaxy selection cut is applied, normalized to similar level as the other curves for qualitative comparison. 
We see that the true galaxy redshift  distribution peaks at a redshift approximately 0.33--0.51 greater than the $\neff$ redshift distribution. 
This is because the high-redshift galaxies are usually poorly measured and contribute less to $\neff$, causing the $\neff$ redshift 
distribution to shift toward lower redshift. We fit the $\neff$ redshift distribution with the following functional form, which is often used to 
characterize the galaxy redshift distribution,
\begin{linenomath*}\begin{equation} 
P(z)=z^{\alpha}exp\left[ -\left( \frac{z}{z_{0}} \right)^{\beta} \right],
\label{eq:z_dist}
\end{equation} \end{linenomath*}
and list the fitted coefficients in the first three rows of \Tref{tab:median_z}. The median redshift $z_{m}$ extracted from the 
curves in \Fref{fig:neff_z} is also listed in \Tref{tab:median_z}. The shape of the distribution is somewhat different from that 
measured in low-redshift galaxies, where $\alpha\approx 2$ and $\beta\approx 1.5$ \citep{1995ApJ...449L.105S}. 

\Fref{fig:neff_z} and \Tref{tab:median_z} show the first attempt to date to quantitatively calculate the redshift 
distribution of $\neff$. This distribution can be used for more realistic Fisher-matrix forecasting and simulations of cosmic shear 
surveys. Previous calculations \citep{1999ApJ...522L..21H, 2004MNRAS.348..897T, 2007MNRAS.381.1018A} adopt the 
raw galaxy redshift distribution for these applications, which could lead to an over-estimation of the constraining power of the 
cosmic shear surveys. 

\begin{table}
  \centering
  \caption{Best-fit coefficients for \Eref{eq:z_dist} that describes the redshift distribution of $\neff$ (the three curves in \Fref{fig:neff_z}). 
  $k$ refers to different galaxy selection cuts (\Eref{eq:k_def}). Also listed is $z_{m}$, the median redshift for the $\neff$ distributions. 
  For comparison, the last column lists the best-fit coefficients and median redshift for the raw galaxy sample $n$ (corresponding to the thick 
  grey band in \Fref{fig:neff_z}). The redshift cut $0.1<z<3$ is not applied here.}
   \begin{tabular}{ ccccc }
       $k$& $2.0$& $1.0$& $0.5$ & $n$ \\ \hline
     $\alpha$ &1.23& 1.24 & 1.28&  1.25\\
     $z_{0}$ & 0.59 & 0.51 & 0.41 &  1.0 \\
     $\beta$ & 1.05 & 1.01 & 0.97&  1.26\\
     $z_{m}$  & 0.89 & 0.83 &0.71& 1.22 \\  
  \end{tabular}
  \label{tab:median_z}  
\end{table}

\section{Effects of other practicalities}
\label{sec:practicalities}

In addition to the considerations discussed above, several practical factors affect $\neff$ at levels that cannot be neglected. 
We roughly estimate the level of these effects. The results of this section are summarized in \Tref{tab:blends}. 

\begin{enumerate}

\vspace{0.1in}
\item[(1)] \noindent \textbf{Blending}
\vspace{0.05in}

\noindent After the selection cut, depending on the galaxy number density, a fraction of the galaxies will be rejected because they 
are blended with other close-by objects. At the low galaxy number density in existing data, this effect is not severe. As a result, 
the effect of blending has not been studied in detail for previous weak lensing analyses. However, given the high galaxy number density 
expected for LSST (\Tref{tab:cut_neff}), the effect can be significant. We estimate below roughly the effect of blending for LSST. A more 
detailed study will be presented in \citet{D13}.

\begin{figure}
 \begin{center}
 \includegraphics[height=2.5in]{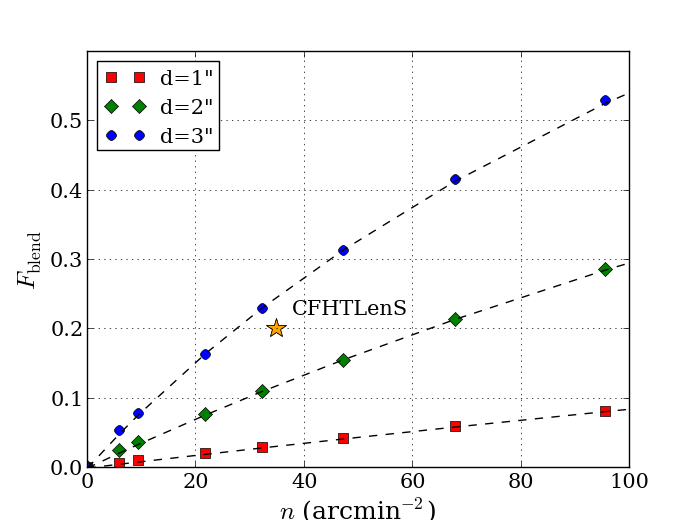} 
 \end{center}
 \caption{Fraction of galaxies that have neighbors with a "center-to-center" distance less than $d$, or blended, as a function of the galaxy number density $n$. The data points 
 are overlaid by the best-fit function (\Eref{eq:fit_blend}, \Tref{tab:blends_fit}) for the three $d$ values with the black dashed curve. The three 
 cases plotted here correspond to conservative ($d=$3"), fiducial ($d=$2"), and optimistic ($d=$1") assumptions for the performance of de-blending 
 algorithms in LSST. 
 The orange star indicates the galaxy number density and fraction of blended galaxies in the CFHTLenS dataset. The de-blending algorithm 
 used in CFHTLenS corresponds to $d\approx2.7$" in our simple model.} 
 \label{fig:neff_blend}
\end{figure}
\begin{table}
  \centering
  \caption{Best-fit coefficients for \Eref{eq:fit_blend} that describes the fraction of galaxies blended as a function of galaxy number density. 
  The fit is evaluated assuming different de-blending algorithms. The three $d$ values correspond to conservative ($d=$3"), fiducial ($d=$2"), 
  and optimistic ($d=$1") assumptions for the performance of de-blending algorithms in LSST.}
   \begin{tabular}{ ccc }
     $d$ & $\eta$ & $\mu$         \\ \hline
     1"    &  0.50 & 1.9 $\times 10^{-3}$  \\ 
     2"    &  0.68  & 5.5 $\times 10^{-3}$\\ 
     3"    &   0.62 &  1.4 $\times 10^{-2}$ \\    
  \end{tabular}
  \label{tab:blends_fit}  
\end{table}

First, we assume a galaxy is ``blended'' when there are other objects with center positions within a radius $d$ of the the galaxy's center, 
where $d$ depends on the capability of the de-blending algorithm assumed. In other words, galaxies with neighbors closer than $d$ will be 
rejected by a certain de-blending algorithm because attempting to de-blend these objects will introduce significant systematic errors on 
their shape measurements. Next, we use the same galaxy catalog introduced in \Sref{sec:sigmaSN} and plot the fraction of blended 
galaxies as a function of galaxy number density in \Fref{fig:neff_blend} for three $d$ values. For de-blending treatments in existing datasets, 
conservative approaches at the level of $d\approx$3" are used (see \Sref{sec:cfhtls}). We expect that when the LSST survey begins, de-blending 
algorithms will be improved, and objects separated by approximately twice the width of the PSF, or $\approx$2", could be properly de-blended. 
$d\approx$1" represents an optimistic case for which the de-blending algorithm is capable of dealing with objects separated by approximately the 
width of the PSF.

The blending fraction $F_{\rm blend}$ in \Fref{fig:neff_blend} can be well described by the functional form
\begin{equation}
F_{\rm blend}(n)=\eta \ln(1+\mu n)\;\;,
\label{eq:fit_blend}
\end{equation}
where $\eta$ and $\mu$ are coefficients to fit for, and $n$ is the galaxy number density. We list the best-fit coefficients in \Tref{tab:blends_fit} 
and overlay the best-fit function to the points in \Fref{fig:neff_blend}. 
In this simplistic blending model, $\neff$ is degraded to $(1-F_{\rm blend})\neff$, given a certain $d$ value. Note that the degradation of 
$\neff$ from blending, or $F_{\rm blend}$, depends on the raw galaxy number density $n$ and not $\neff$. We list the resulting $\neff$ for $d=$2" 
in the second row of \Tref{tab:blends}. 

\vspace{0.1in}
\item[(2)] \noindent \textbf{Masking}
\vspace{0.05in}

\noindent Parts of the images will be masked due to bright stars (generating diffraction spikes, saturated columns, and large 
diffuse halos) and edge effects. Masking can be combined into the $\neff$ calculation or simply accounted for by claiming a 
smaller survey area. In this paper, we choose to account for it in $\neff$ so that the total survey area (18,000 degree$^{2}$) 
is consistent with what one would assume in Fisher-matrix calculations for LSST. 

The fraction of area masked depends heavily on the field observed. For pointings near the Galactic plane, many more bright 
stars will need to be masked. We assume a typical weak lensing field at moderate Galactic latitude 
$b\sim60$ degrees. At this latitude, we estimate approximately 15\% of the image area is masked out. The resulting $\neff$ 
values are listed in the last row in \Tref{tab:blends}. This fraction of masked area is based on typical values obtained in 
existing datasets \citep{2007ApJ...669..714M, 2012ApJ...744..180V, 2012MNRAS.427..146H}. 

Note that a detailed estimation of the masked area for LSST also needs to take into account its unique survey strategy 
(\eg, short-exposure, high-cadence) and instrument design (\eg, highly-segmented CCD sensors). These factors, 
together with the potential improvement in masking techniques will affect the masked area in a non-trivial fashion.
We defer this topic to future studies and adopt the simple estimation described above.

\end{enumerate}

\begin{table}
  \centering
  \caption{Estimation of the $\neff$ accounting for degradation caused by practical effects in data such as blending and masking, 
  for the optimistic ($k=2.0$), fiducial ($k=1.0$) and conservative ($k=0.5$) galaxy selection cuts. These effective number 
  densities are calculated for the last row in \Tref{tab:cut_neff} ($r$- and $i$-band data, combined through joint-fitting). }
   \begin{tabular}{ lccc }
     &\multicolumn{3}{c}{$\neff$} \\ \hline
     $k$ &  2.0 & 1.0 & 0.5         \\ \hline
     raw $\neff$     & 48 & 37   & 24  \\ 
     +blending ($d$=2")  & 36  & 31 & 22 \\    
     +masking (15\%)       & 31  & 26 & 18 \\ 
  \end{tabular}
  \label{tab:blends}  
\end{table}

\section{Case study -- $\neff$ for CFHTLenS}
\label{sec:cfhtls}

To demonstrate the generality of our approach, we perform the same $\neff$ calculation for the 
most recent CFHTLenS dataset. We base the major parameters on the CFHTLenS summary paper 
\citep[][H12]{2012MNRAS.427..146H}, the paper describing the shear measurement pipeline 
\citep{2013MNRAS.429.2858M}, and the CFHTLenS data release paper \citep{2012arXiv1210.8156E}.

First, we point out that in H12, a different definition of the ``effective number density of weak lensing 
galaxies'' is used. To avoid confusion, we will refer to this definition as $\neffh$, where
\begin{linenomath*}\begin{equation} 
  \neffh=\frac{1}{\Omega^{*}}\frac{(\Sigma w_{i}^{*})^2}{\Sigma (w_{i}^{*})^{2}}.
  \label{eq:neffh}
\end{equation} \end{linenomath*} 
\noindent H12 defined $\Omega^{*}$ to be the total area of the survey, \textit{excluding masked regions} and the weighting 
factor $w_{i}^{*}$ is a measure of the shear measurement error, defined as
\begin{linenomath*}\begin{equation} 
w^{*}=\left[ \frac{\sigma_{e}^{2}e_{max}^{2}}{e_{max}^{2}-2\sigma_{e}^{2}} + \sigma_{SN}^{2}\right]^{-1}\;\;,
\end{equation} \end{linenomath*}
\noindent where $\sigma_{e}^{2}$ is the 1D variance in ellipticity of the likelihood surface estimate in 
\textit{lensfit}\footnote{CFHTLenS uses \textit{lensfit} \citep{2007MNRAS.382..315M, 2008MNRAS.390..149K, 
2013MNRAS.429.2858M} as the main shear measurement algorithm.}  
and $e_{max}$ is the maximum allowed ellipticity. In the limit $e_{max}\rightarrow \infty$, the first term in the bracket, 
$(\sigma_{e}^{2}e_{max}^{2})/(e_{max}^{2}-2\sigma_{e}^{2}) $, reduces to $\sigma_{e}^{2}$, which can be associated with 
$\sigma_{m}^2$ in our work. 
In H12, the $\neffh$ for the main lensing sample is calculated to be 11 arcmin$^{-2}$, but $\neff$ 
(as defined in \Eref{eq:neff}) is not calculated. Conceptually, these two quantities measure slightly different 
things: $\neffh$ is a measure of the fraction of all the galaxies used that have measurement noise large 
compared to average measurement noise. $\neffh$ is equal to the raw number density of galaxies selected when all 
the weights are the same and decreases as the distribution of the weights becomes broader. 
On the other hand, $\neff$ is defined specifically to measure the absolute statistical power of a weak lensing dataset and 
is always smaller than the raw number density of galaxies even when all the galaxies have the same nonzero measurement 
noise. In the case of a very conservative cut, where most galaxies selected have low measurement noise and similar 
weighting factor, the $\neffh$ can be close to $\neff$ since they both approach the raw number density of galaxies.

The CFHTLenS dataset covers an area of 154 square degrees, at approximately uniform depth. Each patch 
of sky is imaged 6-7 times with the exposure time in each image being 600 -- 700 s. For simplicity, we assume that all 
fields are imaged 7 times, for 615 s each, making the total exposure time for each galaxy close to the actual total 
exposure time of $\approx$ 4300 seconds. Lensing analyses are performed only in the $i$-band, where the seeing conditions are 
particularly good -- the mean seeing is 0.64" and all images have seeing better than 0.8". We assume the 7 exposures 
have the following equally spaced seeing values: [ 0.48", 0.53", 0.59", 0.64", 0.69", 0.75", 0.80" ]. We also assume that the sky 
background is constant at 20.0 mag/arcsec$^2$. A magnitude cut of $i_{AB}<24.7$ is placed to ensure the shape measurements 
are accurate; we replace the $\sigma_{m}$ cut with this magnitude cut. There is no explicit cut in galaxy size in the shear measurement 
algorithm used in CFHTLenS\footnote{\textit{lensfit} attempts to fit all detected objects with star and galaxy models and then classifies 
them as stars or galaxies according to which model gives a higher posterior probability.}, but since the galaxy models have a prior on 
the scale length set to a minimum of 0.3 pixels, we use that as a measure of the implicit size cut. Using the mean seeing 
0.64"/  (0.187"/pixel) = 3.5 pixels, we have $R_{cut}=(0.3/3.5)^2\approx 0.007$. An additional redshift cut ($0.2<z<1.3$) is 
applied to the galaxy sample used for the lensing analysis to ensure accurate photometric redshift estimates. 
20\% of the galaxies are rejected due to serious blending. 19\% of the survey is masked due to bright stars. A total of 39\% of the 
area is rejected when including the above masking and systematics diagnostics. The main survey quantities of CFHTLenS used 
in this study is summarized in \Tref{tab:cfhtlens}.

The number density of galaxies with shape and redshift measurements, that was used in the CFHTLenS calculation of $\neffh$, is 
$\sim17$ arcmin$^{-2}$. Note that this is much lower than the raw number density of galaxies at $i_{AB}<24.7$ ($\sim 35$ 
arcmin$^{-2}$, estimated from the \catsim catalog). Several effects contribute to the reduced galaxy number density (Miller, Heymans, 
Hildebrandt, private communication): (1) the redshift cut at $0.2<z<1.3$, (2) incompleteness in the photometric redshift data, (3) 
incompleteness in detection due to low surface brightness (4) rejection of large galaxies that are larger than the size of the postage 
stamps ($\sim9$" on a side), and (5) rejection of blended galaxies. Given the expected raw number density of galaxies ($\sim 35$ 
arcmin$^{-2}$) and the final fraction of blended galaxies (20\%), we calculate that the effective blending criteria used in CFHTLenS 
corresponds to $d\approx 2.7$" in \Fref{fig:neff_blend}. 
  
\begin{table}
  \centering
  \caption{Main survey and analysis parameters from CFHTLenS related to the $\neff$ calculation. The asterisks indicate 
  that the values are estimated or approximated from the real parameters in the survey. When applicable, the parameters are 
  specified for the CFHT $i$-band.}
  \begin{tabular}{ ll  }
     \hline
     Total survey area                       &   154 (deg$^2$)     \\             
     Number of exposures*                &     7       \\
     Exposure time per exposure*    &    615 (s) \\
     \multirow{2}{*}{Seeing distribution*}                     &  [ 0.48, 0.53, 0.59, 0.64, \\
                                                            & \hspace{0.1in} 0.69, 0.75, 0.80 ] (")        \\
     Sky background*                          &    20 (mag/arcsec$^2$) \\
     Wavelength range                      &   685 -- 840 (nm) \\
     Pixel scale                                    &    0.187 ("/pixel) \\
     Redshift range                            &    0.2 -- 1.3      \\
     Magnitude range                        &    $i_{AB}<24.7$     \\
     Size cut*                                         &    0.007     \\
     Fraction of galaxies    &   \multirow{2}{*}{20\%}  \\ 
     \hspace{0.1in}rejected due to blending &   \\
     Fraction of area rejected           &   39\%   \\ 
     \hline
  \end{tabular}
   \label{tab:cfhtlens}  
\end{table}

Putting in the above numbers, we calculate that without any blending rejection and masking, we have $\neff \approx$ 12 
arcmin$^{-2}$, a factor of $\sim3$ fewer than the LSST $\neff$ for $r$-band and the fiducial galaxy selection cut. 
Taking into account a 20\% blending rejection, we have 
$\neff \approx10$ arcmin$^{-2}$. As expected for the rather conservative galaxy selection cut, this $\neff$ value is close to the 
$\neffh \approx 11$ arcmin$^{-2}$ calculated in H12.
Finally, taking into account the 39\% rejected area, which is avoided in H12 by considering the un-rejected area 
only, we have $\neff \approx$ 6 arcmin$^{-2}$. This number, together with a shape noise of $\sigma_{SN}\approx0.26$ and 154 
square degree survey area make up a self-consistent set of parameters that quantifies the statistical errors in the CFHTLenS 
cosmic shear measurement. 
Note that one should be careful not to compare the two numbers, $\neff$ and $\neffh$, directly as they measure slightly 
different properties of the galaxy population.


\section{Conclusion}
\label{sec:conclusion}

The effective number density of weak lensing galaxies, or $\neff$, is a measure of the statistical power of a weak lensing survey. 
In this paper, we have conducted a detailed and systematic analysis to calculate $\neff$ for LSST. Our analysis considers all 
major components in a weak lensing pipeline including the galaxy population of interest, the distribution of observing conditions, 
the measurement errors, the galaxy selection procedure, the approach for combining multiple exposures, and blending/masking effects. 
By using realistic simulations, we estimate that with current weak lensing algorithms (the fiducial scenario), $\neff \approx$ 37 arcmin$^{-2}$ 
before considering masking and blending, $\neff \approx$ 31 arcmin$^{-2}$ when rejecting the blended galaxies and $\neff \approx$ 
26 arcmin$^{-2}$ when factoring in a 15\% masked area. This is estimated for LSST after combining all the $r$- and $i$-band data in the 
full 10-year survey on a 18,000 degree$^{2}$ survey area. With improvement in the weak lensing analysis algorithms, we can expect 
(optimistically) $\neff \approx$ 48 arcmin$^{-2}$ before accounting for masking and blending, $\neff \approx$ 36 arcmin$^{-2}$ when blended 
galaxies are rejected and $\neff \approx$ 31 arcmin$^{-2}$ with $15\%$ of the area masked. (\Tref{tab:blends}). 
We have shown quantitatively how improvement in the weak lensing algorithm as well as de-blending/masking techniques can lead to 
large improvements in $\neff$.

Different schemes for combining multiple exposures are 
discussed in this paper. We find that using a co-add method reduces $\neff$ by 7\% compared to a joint-fit method and optimally 
combining data from all six filters could increase $\neff$ by a factor of 1.5 compared to using only the $r$- and $i$-band 
(lensing-optimized) data. We also quantify for the first time the redshift distribution of $\neff$, which has a median redshift 
0.35--0.5 lower than the raw galaxy distribution (\Fref{fig:neff_z}). Finally, we demonstrate how the same methodology can be 
applied to current weak lensing surveys and show that the results are reasonable. 

Now we review and compare our results to the $\neff$ values for LSST that have been estimated in previous studies. 
Before comparing our results with the different studies, it is important to realize two issues regarding the definition of $\neff$. First, 
many other papers define and quote 
differently the ``effective number density of galaxies used for weak lensing measurements''. H12 is an example, described earlier 
(\Eref{eq:neffh}), that has an entirely different definition and underlying meaning of $\neff$. \citet{2006MNRAS.366..101H} and 
\citet{2008A&A...484...67P}, on the other hand, use the number $n_{g}\approx$ 30 arcmin$^{-2}$ in their analyses, where $n_{g}$ is 
the raw number of galaxies instead of the weighted $\neff$. Second, most previous studies do not consider the masked area in the 
$\neff$ calculation. To perform a fair comparison, we will thus compare our $\neff$ estimation \textit{before} masking (second row in 
\Tref{tab:blends}) with other studies. That is, we have $\neff \approx$ 36 arcmin$^{-2}$ (optimistic), $\neff \approx$ 31 arcmin$^{-2}$ 
(fiducial), and $\neff \approx$ 22 arcmin$^{-2}$ (conservative). Under this definition, one needs to specify an additional estimate of 
the masked survey area to fully characterize the statistical power in weak lensing for LSST. 

In \citet{2009arXiv0912.0201L}, $\neff \approx$ 40 arcmin$^{-2}$ was estimated for LSST. This estimate was based on scaling the measurement 
in \citet{2006ApJ...648L.109C} to the LSST depth and includes both $r$- and $i$-band images, with a rough estimate of blending. 
To perform a fair comparison, we note that at the time when the analysis of \citet{2006ApJ...648L.109C} was conducted, it was common 
to use a shear measurement algorithm that is subject to either the ``conservative'' or the ``fiducial'' scenario in this paper. This 
suggests that our estimates in this work, $\neff \approx$ 31 arcmin$^{-2}$ and 36 arcmin$^{-2}$, are relatively lower. We attribute this to the fact 
that first the Clowe study is based on a relatively small area of sky; thus the conclusions may not be general. Second, their blending 
estimation may be overly optimistic. Similarly, an independent group approached the problem by comparing HST and Subaru observations of 
the same field and combining multiple exposures with algorithms that take into account the PSF variation over time \citep[][ Tyson, Dawson and 
Jee, private communication]{2011PASP..123..596J}. They conclude $\neff \approx$ 36 arcmin$^{-2}$ for the full LSST $r$- and $i$-band dataset, 
consistent with our optimistic estimate. In A06, $\neff$ was estimated to be $\approx$ 30 arcmin$^{-2}$ in the pessimistic scenario and $\approx$ 
40 arcmin$^{-2}$ in the optimistic scenario. This is fairly consistent with what we have estimated for the ``fiducial'' and ``optimistic'' scenarios. 
With the detailed study in this paper we have a much better understanding of the origin of $\neff$ and the multiple factors that can affect its level. 
Assumptions about these factors should be specified when stating any $\neff$ estimate. 

Although the focus of this paper is the statistical errors for future lensing surveys, it is important to realize that in many cases systematic 
errors are inevitably coupled with the statistical errors in lensing measurements. In \Sref{sec:gal_selection}, for example, we select 
galaxies with measurement noise lower than a certain threshold for cosmic shear measurement. The main purpose of this selection cut 
is to avoid systematic errors in shear measurements, which become large for faint and small galaxies. Similarly, in \Sref{sec:practicalities} 
we reject galaxies that have close neighbors. This is to avoid shear measurement biases from errors in the de-blending process. 
Finally, incorporating data from filters that are not optimized for lensing does not increase $\neff$ significantly, but multi-color data is 
potentially useful for identifying and quantifying systematic uncertainties. The balance between statistical and systematic errors is the key issue to address when 
designing lensing pipelines in the next generation surveys. After all, it is the combination of the statistical and systematic errors that 
determine the ultimate uncertainties in the cosmological parameters. 


\section*{Acknowledgments}
We thank Gary Bernstein for helpful discussions. We thank Alexie Leauthaud for providing us with the 
COSMOS galaxy catalog. We thank Jon Thaler and Seth Digel for useful comments which have helped 
improve this paper substantially.

Partial funding for MJ and BJ came from the Department of Energy award grant number DE-SC0007901. 
AJC acknowledges support from the Department of Energy award grant number DE-SC0002607. 
LSST project activities 
are supported in part by the National Science Foundation through Governing Cooperative Agreement 0809409 
managed by the Association of Universities for Research in Astronomy (AURA), and the Department of Energy 
under contract DE-AC02-76-SFO0515 with the SLAC National Accelerator Laboratory. Additional LSST funding 
comes from private donations, grants to universities, and in-kind support from LSSTC Institutional Members.

\label{lastpage}


\bsp

\appendix
 
 \section{Calculating the signal-to-noise ratio ($\rm \nu$) and effective size ($\rm R$) of galaxies}
\label{sec:derivation}

For the analysis in this paper, we need to connect the input galaxy and observational parameters with the 
quantities measured from real images. This conversion is essential for calculating the signal-to-noise ratio ($\rm \nu$) and 
effective size ($\rm R$) of galaxies in our analysis.

\subsection{Conversion of galaxy model parameters to observable quantities}
\label{sec:galaxy_param}

For the bulge+disk galaxy model provided by \catsim, we are given the half-light radius of the bulge and the disk 
separately, the total magnitude, and the ratio of flux in the bulge to the total flux (see \Fref{fig:catsim}). The bulge and 
disk are modeled by Sersic profiles with Sersic index $n=4$ and $n=1$ respectively. When measuring the galaxy 
parameters in real data, the galaxy is measured as a whole (bulge+disk). Moreover, sizes of galaxies are often 
measured through moments of the light distribution rather than half-light radius. Here we attempt to connect the two 
conventions. Imagine a galaxy with the following profile:
\begin{linenomath*}\begin{equation} 
I(r)=I_{tot}[f_{b}e^{-(\frac{r}{r_{b}})^{1/4}}+(1-f_{b})e^{-\frac{r}{r_{d}}}]\;\;,
\end{equation} \end{linenomath*}
where the subscript `$b$' and `$d$' indicates the bulge and disk respectively, and $r_{b}$, $r_{d}$ are the 
scale-length for the bulge and disk component. First, we can calculate numerically the relation between the scale lengths 
and the half-light radii, $r_{h,b}$ and $r_{h,d}$, for the individual components:
\begin{linenomath*}\begin{align} 
&  r_{h,b}\approx 3459 r_{b}\;\;, \\ \notag
&  r_{h,d}\approx 1.68 r_{d}\;\;.
\end{align} \end{linenomath*}

Next, we solve analytically for the second-moment radius of these two components: 
\begin{linenomath*}\begin{equation} 
r_{sec, b} \approx 16108 r_{b} \approx 4.66 r_{h,b}\;\;,
\end{equation} \end{linenomath*}
\begin{linenomath*}\begin{equation} 
r_{sec, d} \approx 2.45 r_{d} \approx 1.46 r_{h,d}\;\;,
\end{equation} \end{linenomath*}
where the second-moment radius is defined by
\begin{linenomath*}\begin{equation} 
r_{sec}=\left[ \frac{\int_{0}^{2\pi}\int_{0}^{\infty}r^{2}I(r)rdrd\theta}{\int_{0}^{2\pi}\int_{0}^{\infty}I(r)rdrd\theta} \right]^{0.5} \;\;.
\label{eq:2nd_moment}
\end{equation} \end{linenomath*}
Since the moments are additive component-wise, we have:
\begin{linenomath*}\begin{align} 
\label{eq:total_2nd_moment}
r_{sec}=&\sqrt{f_{b}r_{sec,b}^{2}+(1-f_{b})r_{sec,d}^{2}} \\ \notag
=&\sqrt{f_{b}(4.66 r_{h,b})^{2}+(1-f_{b})(1.46 r_{h,d})^{2}}
\end{align} \end{linenomath*}

\subsection{Connecting input and measured signal-to-noise ratio ($\nu$) and effective size ($\rm R$)}
\label{sec:nu_R}

Now we can calculate \Eref{eq:snr} and \Eref{eq:effR} from the input catalogs as well as the measured images. 

\begin{figure}
 \begin{center}
 \includegraphics[height=2.5in]{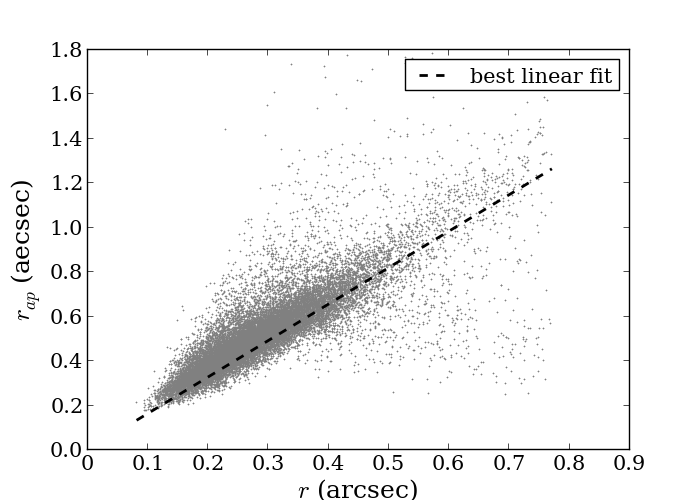} 
 \end{center}
 \caption{Relation between the second-moment radius and the aperture radius as measured from simulations with a measurement 
 noise cut $k=1.0$ (\Eref{eq:k_def}). The relation is approximately linear and can be fitted with the black dashed line with slope 
 $\sim$1.64.} 
 \label{fig:r2_r1}
\end{figure}

From the input galaxy catalog, we first calculate the galaxy's effective size $\rm R$ by finding $r_{\rm gal}$ and $r_{\rm PSF}$ 
separately. We use the second line of \Eref{eq:total_2nd_moment} to calculate $r_{\rm gal}=r_{sec}$, and then calculate $r_{\rm PSF}$ from 
\begin{linenomath*}\begin{equation} 
r_{\rm PSF}=\frac{\sqrt{2}r_{\rm seeing}}{2\sqrt{2\ln2}}\;\;,
\end{equation} \end{linenomath*}
where the `$r_{\rm seeing}$' factor is usually expressed in terms of the 1D full-width-half-maximum (FWHM) size, and the $\sqrt{2}$ in 
the numerator is to convert the 1D quantity to 2D. Next, to calculate the signal-to-noise ratio $\nu$, we first derive the second-moment 
radius of the convolved galaxy image using
\begin{linenomath*}\begin{equation} 
r^2=r_{\rm gal}^2+r_{\rm PSF}^2\;\; .
\end{equation} \end{linenomath*}
We then need the aperture radius $r_{ap}$ for the convolved galaxy. In \citet{1980ApJS...43..305K}, $r_{ap}$ is defined to be
\begin{linenomath*}\begin{equation} 
r_{ap}=2r_{first}=2\frac{\int_{0}^{2\pi}\int_{0}^{\infty}rI^{*}(r)rdrd\theta}{\int_{0}^{2\pi}\int_{0}^{\infty}I^{*}(r)rdrd\theta},
\label{eq:r_ap}
\end{equation} \end{linenomath*}
where $I^{*}(r)$ is the radial profile of the convolved galaxy. Instead of calculating this from the catalog, which is complicated by the 
PSF convolution, we find empirically the relation between the second-moment radius and the aperture radius in typical galaxies in the 
simulations used in \Sref{sec:phosims}. \Fref{fig:r2_r1} shows that for the galaxy sample with measurement noise cut 
$\sigma_{m}<\sigma_{SN}$, we find a simple linear relation between $r$, the convolved galaxy second-moment radius, and $r_{ap}$, 
the aperture radius:
\begin{linenomath*}\begin{equation} 
r_{ap}\approx 1.64 r\;\;.
\end{equation} \end{linenomath*}
Varying the measurement noise cut changes the relation, but for the galaxy sample of interest here, $r_{ap}$ is generally 1.5--2 
times the second-moment radius $r$. In the main analysis of this paper, we use the fiducial cut ($k=1.0$) and assume $r_{ap}\approx1.64 r$. 
In the worse-case scenario where $r_{ap}\approx2 r$ (corresponding to $k=0.5$), we have $\sim$17\% decrease in $\neff$. The source count 
in this aperture is estimated as 90\% of the source count derived by the total source magnitude, while the background count is just the area of 
the aperture times the background flux, which is calculated from \opsim. Both source and background counts will need to be 
multiplied by the throughput of the system in the filter used. We use the average throughput at airmass 1.2 for LSST listed in 
\Tref{tab:throughput}. 

\begin{table}
  \centering
  \caption{Average throughput for LSST across each filter band at airmass 1.2. This includes the throughput of the atmosphere, the 
  optics and the detectors.}
  \begin{tabular}{ ccccccc  }
     filter & $u$ & $g$ & $r$ & $i$ & $z$ & $y$ \\
     \hline
     throughput (\%) &23.5&46.9&51.1&49.0&47.2 &18.8
  \end{tabular}
   \label{tab:throughput}  
\end{table}

From the measured image, we identify the IMCAT output parameters `$\rm r_{g}$'  as the second-moment radius 
(\Eref{eq:2nd_moment}) of the measured objects. The effective radius $\rm R=r_{\rm gal}^2/r_{\rm PSF}^2$ is then calculated from 
\begin{linenomath*}\begin{equation} 
r_{\rm PSF}=\rm r_{g,{star}}\;\;,
\end{equation} \end{linenomath*}
\begin{linenomath*}\begin{equation} 
r_{\rm gal}=\sqrt{\rm r_{g}^{2} - r_{g,{star}}^{2}}\;\;.
\end{equation} \end{linenomath*}

The aperture flux ($S$) and noise ($\sqrt{S+B}$) are outputs of the Source Extractor catalog (`$\rm FLUX\_BEST$' and 
`$\rm FLUXERR\_BEST$'). We follow the convention of \citet{2007ApJS..172..219L} and divide the two parameters to get 
the signal-to-noise ratio ($\rm \nu$) of galaxies. We note that this approach neglects the fact that 
`$\rm FLUXERR\_BEST$' will include the effects of correlated uncertainties in the source size and shape parameters.

\section{Effect of PSF interpolation}
\label{sec:truePSF}

In the main analysis of this 
paper we used a rather idealized PSF model to avoid the PSF estimation problems that can vary significantly from 
pipeline to pipeline. Here, we demonstrate how $\neff$ changes if we consider a more realistic case for PSF 
interpolation.   

PSF interpolation refers to the procedure where we interpolate the PSF model parameters from the stellar positions 
onto the galaxy positions. Conventionally, one would use a smooth low-order polynomial to fit each parameter over 
the field. This interpolation scheme, however, has been shown to be problematic when there are high spatial 
frequency PSF variations from the atmosphere in short exposures \citep{2012MNRAS.421..381H, 
2012MNRAS.427.2572C}. We examine below how this imperfect PSF model degrades $\neff$. 

We generate 1,000 simulated images similar to that described in \Sref{sec:phosims}. In this set of images, we include 
a realistic distribution of stars based on \citet{2008ApJ...684..287I}. We then interpolate the shape parameters of the stellar 
images to the galaxy positions and perform the same PSF correction and shear measurement analysis as before, using 
these interpolated PSF models. The measurement noise surface as a function of $\nu$ and $\rm R$ looks very similar to the 
corresponding interpolated PSF version, but with slightly lower levels. We fix $c$ in \Eref{eq:fit_sigmam} and get $a=2.28$, 
$b=1.1$ when fitted to the measurements. This increase in measurement error yields a very slight decrease in $\neff$ as 
listed in \Tref{tab:neff_reaPSF}. This implies that the actual PSF interpolation method used does not seriously affect $\neff$. 
\begin{table}
  \centering
  \caption{$\neff$ in the case of a more realistic PSF interpolation method, compared to the ideal PSF model used in the 
  main analysis, for the optimistic ($k=2.0$), fiducial ($k=1.0$) and conservative ($k=0.5$) galaxy selection cuts (\Eref{eq:k_def}). 
  The difference in the PSF interpolation scheme causes a small (3--4\%) degradation in $\neff$.}
   \begin{tabular}{ lccc  }
     $k$ &  2.0 & 1.0 & 0.5  \\ \hline
     True PSF & 48 & 37 & 24  \\ 
     Interpolated PSF  & 47   & 36   & 24  \\ 
  \end{tabular} 
  \label{tab:neff_reaPSF}  
\end{table}

\end{document}